\documentclass[pdflatex,sn-mathphys-num]{sn-jnl}

\usepackage{graphicx}%
\usepackage{float}
\usepackage[utf8]{inputenc}
\usepackage[T1]{fontenc}
\usepackage{mathptmx}
\usepackage{etoolbox}

\usepackage{geometry}
\usepackage[normalem]{ulem} 

\usepackage{hyperref}

\usepackage{multirow}%
\usepackage{amsmath,amssymb,amsfonts}%
\usepackage{amsthm}%
\usepackage{mathrsfs}%
\usepackage[title]{appendix}%
\usepackage{xcolor}%
\definecolor{darkgreen}{rgb}{0.0,0.5,0.0}
\usepackage{textcomp}%
\usepackage{manyfoot}%
\usepackage{booktabs}%
\usepackage{algorithm}%
\usepackage{algorithmicx}%
\usepackage{algpseudocode}%
\usepackage{listings}%
\usepackage{booktabs}

\theoremstyle{thmstyleone}%
\theoremstyle{thmstyletwo}%

\theoremstyle{thmstylethree}%

\newcommand{\der}{\text{d}}

\newcommand{\sa}{S^{\alpha}}

\newcommand{\ulen}{l}
\newcommand{\utime}{\tau}
\newcommand{\umass}{m}
\newcommand{\uener}{\epsilon}

\begin{document}

\title[Mesoscale Simulations of Thrombin Activation and Fibrin Formation in Microvascular and \textit{In Vitro} Settings]{Mesoscale Simulations of Thrombin Activation and Fibrin Formation in Microvascular and \textit{In Vitro} Settings}


\author*[1,2]{\fnm{Marina} \sur{Echeverr\'ia Ferrero}}\email{mecheverria@bcamath.org}

\author*[1]{\fnm{Nicolas} \sur{Moreno}}\email{nmoreno@bcamath.org}


\author[1,3,4]{\fnm{Marco} \sur{Ellero}}

\affil[1]{\orgdiv{CFDMS Group}, \orgname{Basque Center for Applied Mathematics }, \orgaddress{\street{Alameda de Mazarredo 14}, \city{Bilbao}, \postcode{48009}, \state{Bizkaia}, \country{Spain}}}

\affil[2]{\orgname{University of the Basque Country UPV/EHU}, \orgaddress{\street{Barrio de Sarriena}, \city{Leioa}, \postcode{48940}, \state{Bizkaia}, \country{Spain}}}

\affil[3]{\orgname{IKERBASQUE, Basque Foundation for Science}, \orgaddress{\street{Mar\'ia D\'iaz de Haro}, \city{Bilbao}, \postcode{48013}, \state{Bizkaia}, \country{Spain}}}

\affil[4]{\orgdiv{Department of Chemical Engineering}, \orgname{Swansea University}, \orgaddress{\city{Swansea}, \postcode{SA1 8EN}, \country{UK}}}


\abstract{Blood coagulation is governed by tightly regulated reaction networks that unfold within a flowing, heterogeneous microvascular environment. Reduced kinetic models of the intrinsic and extrinsic pathways have seen limited \textit{in vitro} validation, and their behavior within spatially resolved flow fields remains largely unexplored. Here, we embed two established reduced networks into a recently proposed mesoscale particle-based framework that resolves fluid momentum transport alongside multispecies advection–diffusion–reaction dynamics. We investigate the initiation phase of coagulation by simulating thrombin formation in microvascular geometries and \textit{in vitro} assays, and we assess the framework’s ability to reproduce thrombin generation curves (TGCs) under physiologically relevant conditions. We further examine how variations in fibrinogen levels —an important determinant of clot structure and a biomarker for inflammation and thrombosis— affect thrombin and fibrin formation. Overall, this study provides a unified computational approach for analysing how biochemical kinetics interact with transport processes, offering insights relevant to thrombosis modeling and blood diagnostics.
}




\maketitle
\section{Introduction}\label{sec1:Intro}
Blood clotting, or coagulation, is a complex process driven by a cascade of chemical reactions and cellular interactions within the bloodstream. It is broadly categorized into three stages: initiation, amplification, and propagation \cite{hoffman2001cell}. A critical outcome of this process is the formation of a fibrin network, which stabilizes the clot and prevents excessive bleeding. This involves the enzymatic conversion of soluble fibrinogen into insoluble fibrin by thrombin, a key protease generated through a sequence of proteolytic reactions involving clotting factors such as prothrombin and Factor X. Notably, these reactions occur primarily on the surface of activated platelets, which provide a localized environment for the clotting factors to interact.

\begin{figure}[!htbp]
    \centering
    \includegraphics[width=.67\textwidth]{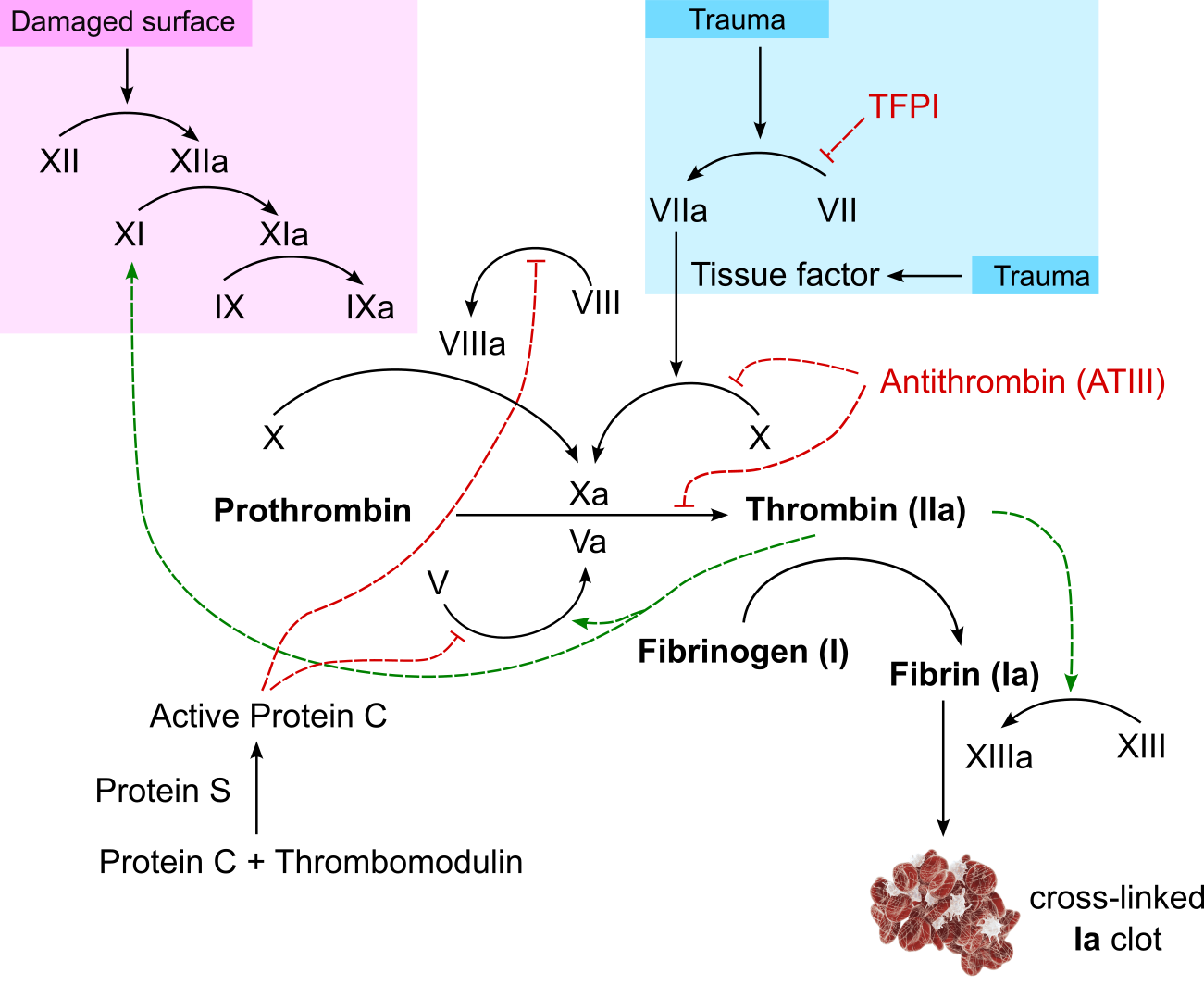}
    \caption{Flowchart of the coagulation network during the initiation phase. The extrinsic pathway (blue) is triggered by tissue trauma, while the intrinsic pathway (pink) begins with Factor XII activation. Both converge at Factor X activation to Factor Xa, leading to thrombin (IIa) generation. Thrombin cleaves fibrinogen (I) into fibrin (Ia) and initiates three positive feedback loops (Factors V, VIII, XI; green dashed lines). Major inhibitors, including TFPI and Antithrombin, are shown in red.}
    \label{fig:RD_Intro_SystemsBioCoagulation}
\end{figure}

While the initiation phase of coagulation is relevant across vessels of all sizes, its dynamics take on particular significance in the microvasculature \cite{falati2003,brass2016}. Here, smaller vessel dimensions, slower flow rates, and higher surface-to-volume ratios amplify the localized effects of tissue factor (TF) exposure and platelet activation. These features create conditions where clotting must be tightly regulated to balance hemostasis and thrombosis. During the initial stages of clotting (Fig. \ref{fig:RD_Intro_SystemsBioCoagulation}), TF, exposed due to vessel injury or inflammation, triggers the extrinsic pathway, leading to Factor VIIa activation and subsequent Factor X activation. Similarly, the intrinsic pathway, initiated by Factor XII activation on collagen or negatively charged surfaces, also converges at Factor X activation \cite{davie1991,hoffman2001cell}. Factor Xa cleaves prothrombin (II) to thrombin (IIa), which not only catalyzes fibrin (Ia) formation but also amplifies the cascade through feedback loops.

Mathematical and computational models of coagulation have expanded greatly, with a large number of publications indexed in PubMed since the 1970s \cite{pubmed2025}. These models aim to capture the interplay of coagulation factors, platelet activity, fluid flow, and vessel wall interactions \cite{fogelson2006,fogelson2008,anand2003model,chatterjee2010systems}. Yet, reproducing the full multiscale physiology remains challenging \cite{leiderman2021art, xu2009study}, as it requires coupling reaction kinetics with diffusion, advection, and cellular processes. Early ODE-based kinetic models described thrombin generation but neglected spatial transport \cite{hockin2002model,brummel1999}. Reaction–diffusion (RD) models later addressed this by adding diffusion, though at the cost of greater computational load, reduced reproducibility, and difficulty in validation \cite{kuharsky2001surface,bodnar2014}. Numerical stability can also be compromised by stiff reaction terms. Reduced models based on physiologically motivated simplifications have been proposed, retaining essential dynamics while limiting parameter complexity \cite{anand2022,bodnar2014}.

To better capture \textit{in vivo} conditions, RD models were further extended to include flow, resulting in advection–diffusion–reaction (ADR) frameworks \cite{xu2009study,leiderman2011grow}. These models enable the simulation of clot formation under physiologically relevant flow regimes by accounting for the influence of transport \cite{kuharsky2001surface,anand2006} and shear forces \cite{kuharsky2001surface,hathcock2006flow,flamm2012multiscale}. Yet, reproducing the full multiscale physiology remains challenging \cite{owen2024}, as it requires coupling reaction kinetics with diffusion, advection, and cellular processes.

Particle-based methods have played an important role in advancing spatially resolved models of thrombosis, complementing continuum and hybrid approaches. Reactive DPD frameworks have been used to explore fibrin polymerization dynamics \cite{yesudasan2018fibrin}, while hybrid particle–continuum models have captured flow–thrombus interactions at mesoscale resolution \cite{mukherjee2018,teeraratkul2021}. Alongside these developments, the large biochemical scope of the coagulation cascade has motivated substantial efforts toward subsystem or reduced-order formulations that isolate key drivers of thrombin generation. Early kinetic models, including those by Hockin et al. \cite{hockin2002model}, Anand et al. \cite{anand2003model}, and Chatterjee et al. \cite{chatterjee2010systems}, provided key insights while keeping parameter counts tractable. Building on these foundations, Fogelson and colleagues incorporated flow and surface effects \cite{kuharsky2001surface,fogelson2006,fogelson2008}, highlighting the importance of spatial and hemodynamic regulation in clotting. More recent reduction strategies --such as multistep model-compression methods \cite{chen2025multistep}, the thrombin-focused schemes of Chen and Diamond \cite{chen2019reduced}, patient-specific calibrations by Ratto et al. \cite{ratto2021patient}, and an SPH formulation in which particles transition from fluid to solid based on a reduced subset of coagulation species and mechanical triggers-- together highlight the continuing need for models that capture essential biology while remaining easy to interpret \cite{monteleone2023}.

In parallel, increasingly realistic multiphysics frameworks have integrated biochemical kinetics with image-based platelet aggregation, reaction–diffusion descriptions of fibrin polymerization, and fluid-mechanical characterizations of shear-dependent clot deformation. These multiscale approaches \cite{bouchnita2019multiscale,ye2020key,shankar2022three,li2022multiphysics} provide a comprehensive view of thrombosis mechanisms, but often rely on extensive reaction networks or tightly coupled modules whose calibration remains challenging. Even focused sensitivity analyses, such as the 88-equation system examined by Stobb et al. \cite{stobb2024}, illustrate the inherent trade-off between biological realism and computational tractability. Despite this progress, important questions remain regarding how flow, geometry, and spatially evolving biochemical fields jointly regulate the initiation phase of clot formation; existing ADR models typically rely on continuum formulations that do not capture mesoscale hydrodynamics, while particle- or agent-based approaches often incorporate only limited reaction networks \cite{owen2024}.

The primary aim of this work is to assess how established reduced kinetic models of the intrinsic \cite{ratto2021patient} and extrinsic \cite{chen2019reduced} pathways behave when embedded in a spatially resolved mesoscale flow framework capable of resolving both transport and reaction dynamics. To this end, we couple these validated reaction networks with a thermodynamically consistent particle-based method \cite{echeverria2025smoothed} that represents fluid momentum, advection–diffusion of macromolecular species, and their nonlinear interactions.

Rather than developing new kinetic schemes, our goal is to systematically examine how these models interact with transport mechanisms under controlled hemodynamic and biochemical conditions. We explore thrombin and fibrin formation in vessel-like microfluidic geometries as well as static \textit{in vitro} domains, using thrombin generation curves (TGCs) \cite{hemker2003,deLaat2020} as a benchmark for initiation, amplification, and termination dynamics. We further illustrate the flexibility of the framework by analyzing selected biochemical perturbations, including variations in fibrinogen levels, critical in pathological conditions such as inflammation, cardiovascular disease, and COVID-19 coagulopathy \cite{kangro2022fibrinogen}. Finally, we discuss the main biological insights and implications of our findings and their relevance for multiscale thrombosis modeling and blood diagnostics.

\section{Computational model}\label{sec:compumodel}

\subsection{Compositional smoothed dissipative particle dynamics}\label{subsec:sdpd-formulation}
We model blood flow and species transport using a compositional extension of smoothed dissipative particle dynamics (SDPD) developed for mesoscale advection–diffusion–reaction problems \cite{echeverria2025smoothed}. This framework builds on the hydrodynamic formulation of standard SDPD \cite{Espanol2003,Ellero2003,Petsev2016} and augments it by evolving particle-based concentrations of multiple biochemical species.

At the continuum level, SDPD is consistent with the compressible Navier–Stokes and mass-balance equations:
\begin{equation}
\frac{\mathrm{d} \rho}{\mathrm{d} t} = - \rho \nabla \cdot \mathbf{v},
\end{equation}
\begin{equation}
\rho \frac{\mathrm{d} \mathbf{v}}{\mathrm{d} t} + \nabla p - \eta \nabla^2 \mathbf{v} - \left(\zeta + \frac{\eta}{\mathcal{D}}\right) \nabla (\nabla \cdot \mathbf{v}) = 0,
\qquad \mathcal{D}=2,3,
\end{equation}
where $\rho$ denotes the density, $\mathbf{v}$ the velocity, $p$ the pressure, and $\eta$ and $\zeta$ the shear and bulk viscosities.

Species transport is described by the advection–diffusion–reaction (ADR) equation. Each chemical constituent $\alpha$ is represented by a concentration field $C^\alpha(\mathbf{x},t)$, which is advected with the fluid and evolves due to prescribed diffusivity and local reaction kinetics:
\begin{equation}
\frac{\mathrm{d} C^\alpha}{\mathrm{d} t}
= - C^\alpha (\nabla \cdot \mathbf{v}) + \nabla \cdot \left( \frac{T C^\alpha}{\xi^\alpha} \nabla \frac{\mu^\alpha}{T} \right)
+ S^\alpha .
\label{eq:compBalance} 
\end{equation}

The second term describes diffusive transport in a thermodynamically consistent form; for dilute species, the expression $\nabla \cdot \big( {(T C^\alpha)}/{\xi^\alpha},\nabla (\mu^\alpha/T) \big)$ reduces to the classical Fickian form $\nabla \cdot ( D^\alpha \nabla C^\alpha )$, with an effective diffusivity $D^\alpha = T/\xi^\alpha$. Here, $T$ is the temperature, $\mu^\alpha$ the chemical potential, $\xi^\alpha$ a mobility coefficient that sets the diffusivity. Finally, the third term $S^\alpha$ accounts for biochemical reactions.

In the particle discretisation, the fluid is represented by SDPD particles that carry mass, momentum, thermodynamic state, and concentrations $C_i^\alpha$ evolve according to a particle discretisation of the continuum ADR equation above, involving pairwise diffusive fluxes between neighbouring particles and local reaction terms. This compositional extension allows species-specific diffusivities to be prescribed while retaining the thermodynamic consistency of the underlying fluctuating hydrodynamics model, making it suitable for simulating macromolecular transport and reaction dynamics in micro/meso-vascular flows. The complete discrete forms, including the definition of the pairwise diffusivity, interpolation kernels, and concentration update, are provided in Appendix \ref{app:subsec:SDPD}.

We adopt a system of dimensionless (reduced) units by defining characteristic scales for length ($\ulen$), mass ($\umass$), and energy ($\uener$). These scales yield a natural unit of time, $\utime = \ulen \sqrt{\umass/\uener}$, and all quantities are expressed relative to them. For example, a domain length $L_x = 50$ corresponds to $50\ulen$, and an interparticle spacing $dx = 0.2$ provides a resolution of $0.2\ulen$. Transport and reaction parameters are likewise nondimensionalized: diffusion coefficients in $\ulen^2/\utime$, reaction rates in $1/\utime$, and velocities in $\ulen/\utime$. This framework ensures internal consistency and facilitates comparison across test cases.

To characterize transport regimes, we express the results using standard dimensionless numbers. The Reynolds number, $\mathrm{Re} = \rho u \ulen / \eta = u \ulen / \nu$, measures the ratio of inertial to viscous forces, where $\rho$ is the fluid density, $u$ the characteristic velocity, $\eta$ the dynamic viscosity, and $\nu$ the kinematic viscosity. The Peclet number, $\mathrm{Pe} = u \ulen / D$, quantifies the relative importance of advection and diffusion, while the Schmidt number, $\mathrm{Sc} = \nu / D$, compares momentum and mass transport. Finally, the Damk\"ohler number, $\mathrm{Da} = \ulen^2 / (D \tau_\mathrm{r})$, relates diffusion and reaction timescales, with $\tau_\mathrm{r}$ the characteristic reaction time. Additional methodological details, including stability criteria, are summarized in Appendix \ref{app:sdpd:properties}.

In this work, we focus on the microvascular regime, characterized by vessel diameters of 5–100 $\mu$m, low Reynolds numbers (Re $\lesssim 1$), and moderate-to-high Peclet numbers (Pe $\sim 10^1{-}10^4$). Such conditions typify arterioles and venules and are often reproduced in flow-based \textit{in vitro} assays of coagulation and thrombus formation \cite{mangin2021vitro,nagy2017,popel2005}. This regime is physiologically and experimentally relevant and well suited for mesoscale modeling, where transport and reaction dynamics are strongly coupled. The corresponding parameter estimates for our simulations and detailed description range of Pe and Re simulated are provided in Appendix \ref{app:subsec:RePe}. The choice of characteristic scales and non-dimensional parameters ensures that the simulated flow remains within the physiological range of arteriolar conditions. The ADR–SDPD scheme is implemented in the open-source LAMMPS \cite{Thompson2022lammps} (stable release: 29 Oct 2020), which offers excellent scalability for large simulations and compatibility with existing modules for molecular and rigid-body systems.

\subsection{Mathematical models for the coagulation pathways}\label{subsec:coagulmodels}

We next introduce the sets of ordinary differential equations (ODEs) used to describe the biochemical kinetics of the coagulation pathways. These equations correspond to the source term $\sa$ in the compositional balance equation \eqref{eq:compBalance}, and provide the kinetic component of the SDPD framework. To simplify the notation, the compositions of each species, $C^k$ are denoted by the abbreviation of the species (e.g. $C^\mathrm{IIa} = \mathrm{IIa}$).

\subsubsection{\label{subsubsec:intrinsic}\textit{Intrinsic} coagulation pathway}

Ratto et al.~\cite{ratto2021patient} reduced the intrinsic coagulation pathway to a minimal system of three species: activated Factor X (Xa), prothrombin (II), and thrombin (IIa). Their dynamics are described by nine rate constants that capture initiation (\(k_1\)), propagation (positive terms), and termination (negative terms). The resulting system of ODEs reads
\begin{subequations}
\begin{align}
&\frac{d \mathrm{Xa}}{d t}  = \left(k_1+k_2 \mathrm{IIa}+k_3 (\mathrm{IIa})^2\right)\left(\mathrm{X}^0-\mathrm{Xa}\right)-k_4 \mathrm{Xa}, \label{eq:RD_Rattoa}\\
&\frac{d \mathrm{II}}{d t} =  - \left(k_5 \mathrm{Xa} + k_6 \mathrm{IIa} + k_7 (\mathrm{IIa})^2 + k_8 (\mathrm{IIa})^3\right)\ \mathrm{II}, \label{eq:RD_Rattob}\\
&\frac{d \mathrm{IIa}}{d t} = \left(k_5 \mathrm{Xa}+k_6 \mathrm{IIa}+k_7 (\mathrm{IIa})^2+k_8 (\mathrm{IIa})^3 \right)\ \mathrm{II}-k_9 \mathrm{IIa}, \label{eq:RD_Rattoc}\\
&\frac{d \mathrm{fbg}}{d t} = -\alpha_1 \mathrm{IIa}\, \mathrm{fbg}, \label{eq:RD_Rattod}\\
&\frac{d \mathrm{Fbn}}{d t} =  \alpha_1 \mathrm{IIa} \,\mathrm{fbg}-\alpha_2 \mathrm{Fbn}, \label{eq:RD_Rattoe}\\
&\frac{d \mathrm{Fbn}_P}{d t} =  \alpha_2 \mathrm{Fbn}. \label{eq:RD_Rattof}
\end{align}
\label{eq:RD_Ratto}
\end{subequations}
Equations \eqref{eq:RD_Rattoa}-\eqref{eq:RD_Rattoc} constitute the minimal system employed in preliminary tests. To capture fibrin formation, the model is extended with Eqs. \eqref{eq:RD_Rattod}-\eqref{eq:RD_Rattof}, where fibrinogen (\(\mathrm{fbg}\)) is cleaved by thrombin (IIa) to fibrin (\(\mathrm{Fbn}\)), which then polymerizes into stable fibrin (\(\mathrm{Fbn}_P\)), as a polymer which does not undergo transport.  In  Appendix \ref{app:subsubsec:ChenDiEqns}, we provide a breakdown of the rate constants and parameters for reaction (\(k_1\)–\(k_9\), \(\alpha_1\), \(\alpha_2\)), and the diffusivities $D^\alpha$ of the species.

\subsubsection{\label{subsubsec:extrinsic}\textit{Extrinsic} coagulation pathway}
We adopt the extrinsic pathway model of Chen and Diamond \cite{chen2019reduced}, which captures thrombin and fibrin formation triggered by TF. The system tracks eight chemical species: TF, Xa, IXa, XIa, fibrin (Fbn), free thrombin (IIa), and IIa bound to Fbn at either weak (${}^{\mathrm{E}}\mathrm{S}$) or strong (${}^{\gamma}\mathrm{S}$) sites. Their dynamics are described by the ODE system in Eqs. \eqref{eq:ChenDi}, with fibrin availability determining binding capacity (${}^{\mathrm{E}}\theta_{\text{total}} = 1.6\cdot\text{Fbn}, \ { }^{\gamma}\theta_{\text{total}} = 0.3\cdot\text{Fbn}$).
\begin{subequations}\label{eq:ChenDi}
\begin{align}
    \dfrac{d\mathrm{TF}}{dt} &= -k_{i,\mathrm{TF}} \cdot \mathrm{TF},\\
    \dfrac{d\mathrm{Xa}}{dt} &= a_1 \cdot \mathrm{TF} + a_3 \cdot \mathrm{IXa} - k_i \cdot \mathrm{Xa},\\
    \dfrac{d\mathrm{IXa}}{dt} &= a_2 \cdot \mathrm{TF} + a_7 \cdot \mathrm{XIa} - {k_i} \cdot \mathrm{IXa}, \\
    \dfrac{d\mathrm{XIa}}{dt}  &= \eta_6 \cdot a_6 \cdot \mathrm{IIa} - k_{\text{elute}} \cdot \mathrm{XIa} - k_i \cdot \mathrm{XIa}, \\
    \dfrac{d\mathrm{Fbn}}{dt} &= \eta_5 \cdot a_5 \cdot \mathrm{IIa},\\
    \dfrac{d\mathrm{IIa}}{dt} &= \eta_4 \cdot a_4 \cdot \mathrm{Xa} 
    - \left( {d^{\mathrm{E}}\mathrm{S}}/{dt} + {d^{\gamma}\mathrm{S}}/{dt} \right) \nonumber \\
    &\quad - k_{\text{elute}} \cdot \mathrm{IIa} + k_i \cdot \mathrm{IIa},\\
    \dfrac{d^{\mathrm{E}}\mathrm{S}}{dt} &= {}^{\mathrm{E}}k_{\mathrm{f}} \cdot \mathrm{IIa} \cdot ({}^{\mathrm{E}}\theta_{\text{total}} - {}^{\mathrm{E}}\mathrm{S}) - {}^{\mathrm{E}}k_{\mathrm{r}} \cdot {}^{\mathrm{E}}\mathrm{S},\\
    \dfrac{d^{\gamma}\mathrm{S}}{dt}  &= {}^{\gamma}k_{\mathrm{f}} \cdot \mathrm{IIa} \cdot ({}^{\gamma}\theta_{\text{total}} - {}^{\gamma}\mathrm{S}) - {}^{\gamma}k_{\mathrm{r}} \cdot {}^{\gamma}\mathrm{S}.
\end{align}
\end{subequations}  
Reaction rates ($\alpha_i$), binding/unbinding constants, inhibition  ($k_i$), and tuning parameters ($\eta_i$) are listed in Appendix \ref{app:subsubsec:ChenDiEqns}.

\subsection{\textit{In vivo} and \textit{in vitro} domain definitions}\label{subsec:domain-def}
Simulations were first performed in homogeneous domains without flow or geometric features, providing a simple setting to test the reaction kinetics in \eqref{eq:RD_Ratto} and \eqref{eq:ChenDi} against reference assays. We then extended to heterogeneous domains including localized injury sites, wall-bound particles, and capped regions, which mimic, in simplified form, vascular or experimental conditions where coagulation is locally initiated. In this way, homogeneous cases served as a baseline for model validation, while heterogeneous domains captured the interplay between flow, transport, and localized activation.

\begin{figure}[htbp]
\centering
\includegraphics[width=.95\linewidth]{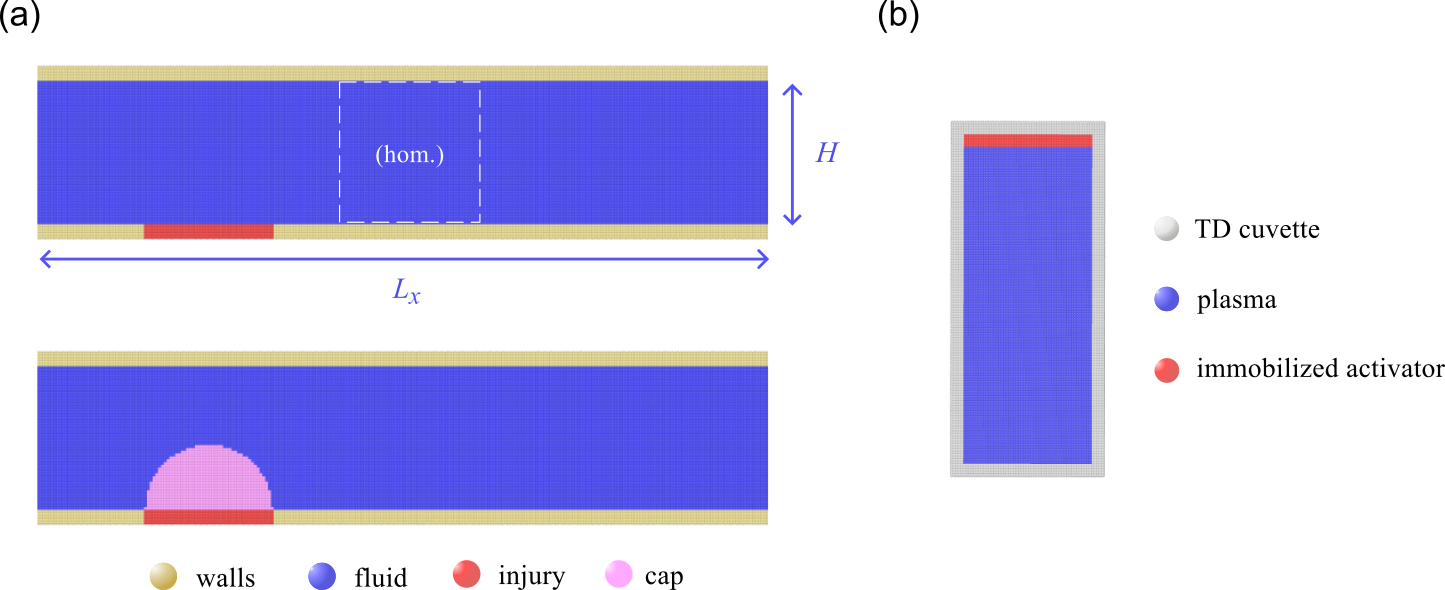}
\caption{Simulation domains for \textit{in vivo} and \textit{in vitro} studies. 
(a) Two rectangular channel domains ($L_x=$55, $H=$10) for \textit{in vivo} simulations. The dashed outline marks the homogeneous (hom.) domain. The heterogeneous domain includes ``injury'' particles (red) to trigger coagulation; a semicircular cap (pink) represents quiescent particle clusters. 13,260 is the total count of fluid particles (blue).
(b) Cuvette for \textit{in vitro} tests, with 6,426 fluid particles; TF (red) initiates coagulation, static cuvette walls (gray).}
\label{fig:ADR_domains}
\end{figure}

Within the heterogeneous group, two main configurations were distinguished.
The first, referred to as \textit{in vivo} domains, consisted of rectangular channels (Fig. \ref{fig:ADR_domains}a) where transport was governed by convection and diffusion, characterized by the Re and Pe ranges introduced in  Section~\ref{subsec:sdpd-formulation}. The computational domain length and height in the flow direction were $L_x=55\ulen$ and $H=10\ulen$, respectively, leading to a systematic exploration of vessel diameters on the range 5–100~$\mu$m,  (Appendix~\ref{app:subsec:RePe}). Fluid particles (blue) represent the plasma phase. Coagulation is initiated by a set of ``injury'' particles (red) localized in the bottom vessel wall. Quiescent particle clusters are represented as a semicircular cap (pink) to mimic resting platelets or microaggregates. The dashed outline in the schematic highlights the homogeneous domain.

The second configuration, referred to as \textit{in vitro} domains, reproduces coagulation assays under quiescent conditions. In particular, we modeled a cuvette geometry (Fig.~\ref{fig:ADR_domains}b) consistent with the Thrombodynamics assay \cite{sinauridze2018thrombodynamics}, which provides spatially resolved measurements of clot initiation and propagation. Here, the SDPD model was scaled to macroscopic cuvette dimensions (centimeter range) and initialized under quiescent conditions to reproduce the diffusion-driven propagation of the fibrin front.
\subsubsection{Boundary and source term conditions}\label{subsubsec:BCs}

Boundary conditions were defined according to the domain type. For particle motion, no-slip conditions were imposed at solid walls, and periodic boundaries were applied elsewhere \cite{Bian2015}. In the \textit{in vivo} channel simulations, prothrombin (II) was continuously supplied at the inlet at either physiological concentrations (see Table \ref{tab:RD_Ratto_modelparameters} in Appendix \ref{app:subsubsec:Ratto}), while at the outlet we imposed homogeneous Neumann conditions, $\mathbf{n}\cdot\nabla C^\alpha = 0$, for the composition of all species. A flux of Factor Xa is applied on particles above the injury-type region to trigger coagulation via the intrinsic pathway, as $\text{flux} = \gamma (C^{\mathrm{X^0}} - C^\mathrm{Xa,avg})$ with $\gamma$ and $C^{\mathrm{X^0}}$ found in Table \ref{tab:RD_Ratto_modelparameters}, and $C^\mathrm{Xa,avg}$ is the dynamically computed average Xa in the fluid only. For semicircular caps, the flux scales with cap area. To reproduce physiologically relevant Pe and Re numbers in the microvasculature, we applied a body force to drive the flow and we adjust the fluid viscosity $\eta$ and the velocity flow $u$ to vary Pe and Re (see Appendix \ref{app:sdpd:properties}).

In the \textit{in vitro} cuvette simulations, coagulation was instead initiated by TF immobilized at the wall under quiescent conditions, with no-flux boundaries applied elsewhere. In the \textit{in vitro} cuvette simulations, TF was immobilized at the wall to initiate coagulation under quiescent conditions. The red region in Fig.~\ref{fig:ADR_domains}b represents the activator surface, assigned a nonzero TF concentration to locally initiate coagulation. No-flux conditions were applied at the remaining boundaries of the domain.

\section{Results}

In the following, we present our results in increasing order of complexity. We begin by calibrating the reduced coagulation models in homogeneous domains to validate the SDPD formulation against ODE-based references. Building on this, we extend the simulations to \textit{in vivo}-like conditions under flow, where we assess the influence of transport parameters, the size of the injury, and variations in fibrinogen concentration on thrombin generation and clot formation. Finally, we performed \textit{in vitro}-inspired simulations emulating experimental assays such as Thrombodynamics \cite{sinauridze2018thrombodynamics}, highlighting the role of diffusion in reaction kinetics under controlled laboratory conditions. Time is made non-dimensional as $t^* = t / t_{\mathrm{ref}}$, where  $t_{\mathrm{ref}}=1$s. Whereas the concentrations are normalized as $C^*=C^\alpha / C_\mathrm{ref}$, with $C_\mathrm{ref}=1~\mu\mathrm{M}$ being a typical inlet fibrinogen concentration ($[\mathrm{fbg}]_{\mathrm{ref}} = 1\ \mu\mathrm{M}$), such that ($C^*=C / C_{\mathrm{ref}}$). This normalization is applied throughout all figures unless otherwise stated.

\subsection{Calibration of the coagulation models}\label{subsec:calibcoagulmodels}

\begin{figure}[!htbp]
\centering
\includegraphics[width=.67\textwidth]{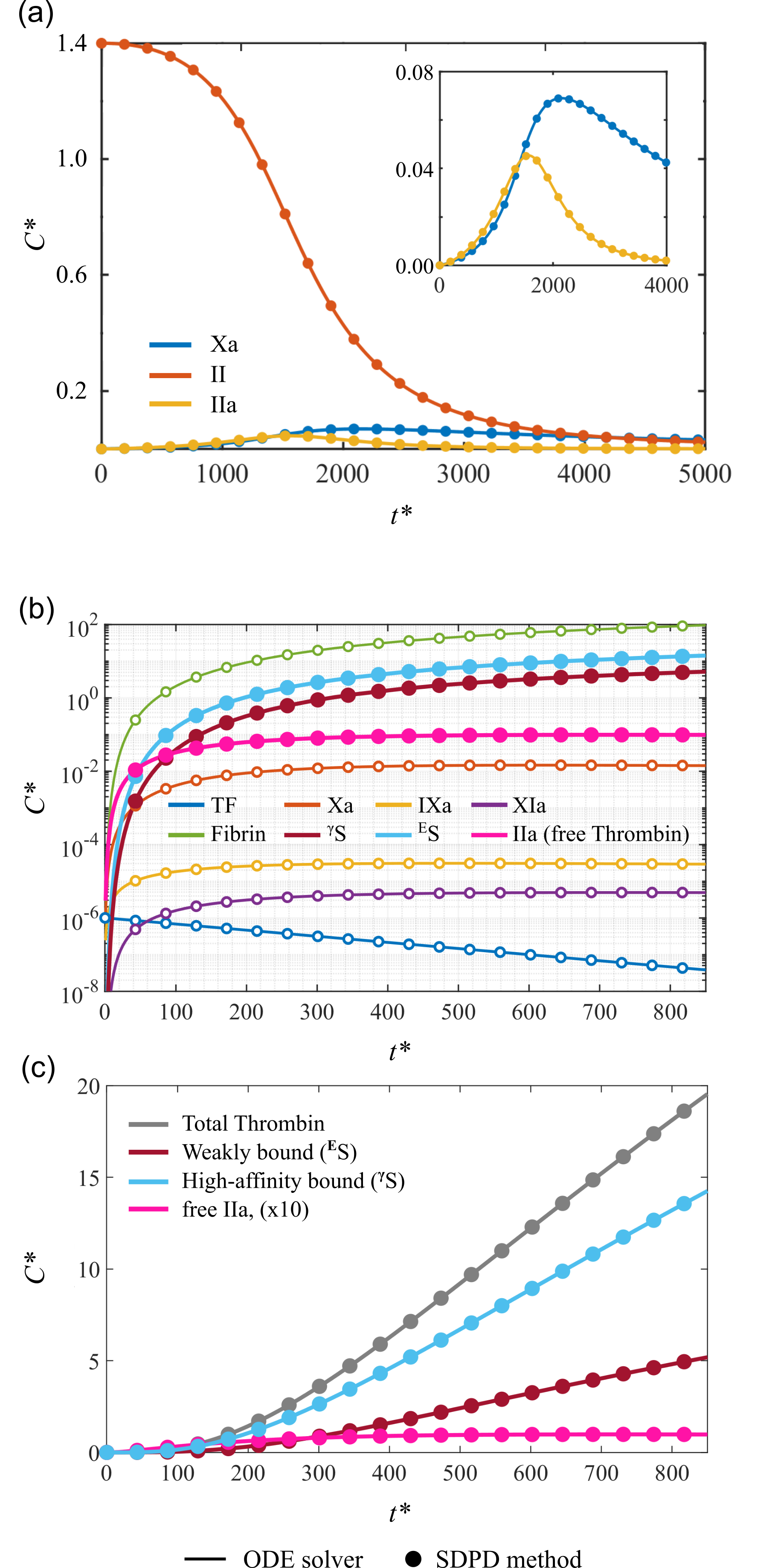}
\caption{(a) \emph{Intrinsic} pathway: Evolution of mean concentration profiles for Factor Xa, prothrombin (II), and thrombin (IIa). Solid lines show MATLAB \texttt{ode15s} results, and circles denote SDPD simulations; the inset highlights Xa and IIa peaks. (b,c) \emph{Extrinsic} pathway: Evolution of mean concentrations in the reduced extrinsic network \cite{chen2019reduced}. Panel (b) shows all species (log scale), and (c) focuses on thrombin forms: strongly bound (${}^{\gamma}\mathrm{S}$), weakly bound (${}^{\mathrm{E}}\mathrm{S}$), free, and total.}
\label{fig:RD_Coagulationtest}
\end{figure}

Simulations to test our SDPD formulation \eqref{eq:compBalance} were first conducted in a homogeneous square domain (dashed region in Fig. \ref{fig:ADR_domains}) for initial calibration and performance assessment. We used the ODE systems described in Section \ref{subsec:coagulmodels} for the intrinsic (Ratto et al. \cite{ratto2021patient}) and extrinsic (Chen and Diamond \cite{chen2019reduced}) pathways. Fig. \ref{fig:RD_Coagulationtest}a shows the concentration profiles of key species in the intrinsic pathway, including Factor Xa, prothrombin (II), and thrombin (IIa), averaged over all fluid particles. Averaged concentration profiles for the species involved in the extrinsic pathway, together with the temporal evolution of IIa, both in its individual forms and total concentration, are shown in Figs. \ref{fig:RD_Coagulationtest}b and \ref{fig:RD_Coagulationtest}c, respectively.

The SDPD simulations successfully reproduce the trends predicted by the ODE systems \eqref{eq:RD_Ratto} and \eqref{eq:ChenDi}, supporting the robustness of our framework for modeling both coagulation pathways. Preliminary tests using the intrinsic pathway model \cite{ratto2021patient} under homogeneous conditions (see Appendix, Fig. \ref{fig:RD_Ratto_checkrates_TGCsArea}) showed negligible differences between simulations with uniform and species-specific diffusivities. This indicates that, under such conditions, coagulation kinetics dominate over diffusive transport (Da $\gg$ 1), and the system is expected to remain primarily reaction-controlled even when spatial transport is introduced in the channel configuration.
\subsection{\textit{In vivo} results}\label{subsec:Resultsinvivo}

We analyzed blood coagulation in platelet-free plasma under flow conditions using system \eqref{eq:RD_Ratto} describing the reduced intrinsic pathway. The results are reported as peak thrombin values and local outlet-averaged thrombin concentrations, which can be directly compared to TGCs as standard benchmarks for the temporal evolution of thrombin. From this point forward, prothrombin and thrombin will be referred to as II and IIa, respectively.

\subsubsection{Effect of P\'eclet number on blood coagulation}\label{subsubsec:Resultsinvivo-Pecletvary}

To assess the role of transport in coagulation dynamics, we examined channel domains at low Re (0.02) while varying Pe (10, 70), consistent with the microvascular regimes introduced in Section \ref{sec1:Intro}. The transport–reaction system comprised Factor Xa, II and IIa, coupled through the kinetic scheme in \eqref{eq:RD_Rattoa}–\eqref{eq:RD_Rattoc} and subject to advection–diffusion.

\begin{figure}[!htbp]
\centering
\includegraphics[width=1
\linewidth]{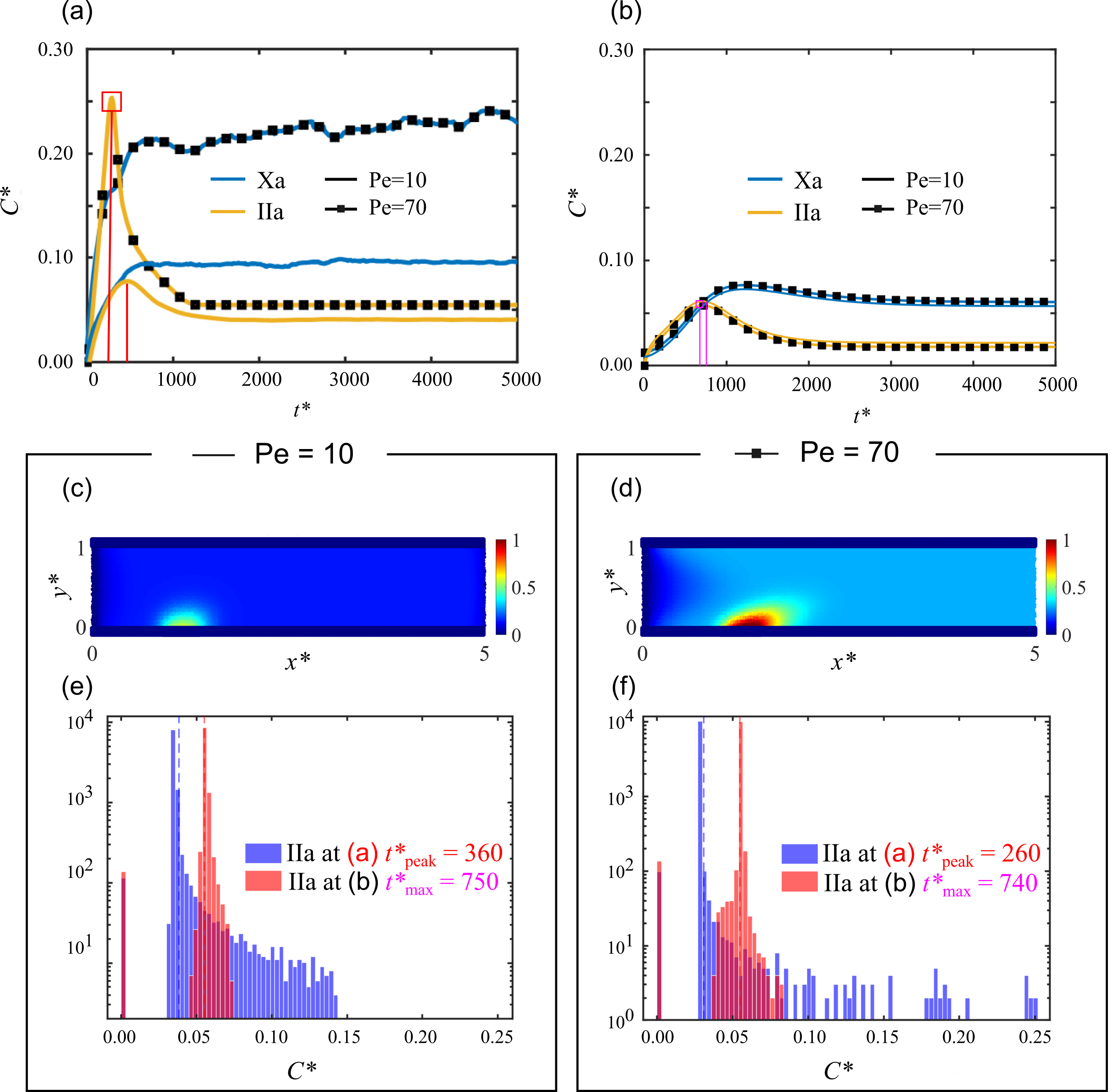}
\caption{Influence of Pe (10,70) under low Re=0.02.
(a) Maximum concentrations of species Xa (blue) and IIa (yellow) over time, with solid lines for Pe$=$10 and filled squares for Pe$=$70. The red vertical lines highlight the time of maximal absolute IIa peak occurrence, $t^*_{\mathrm{peak}}$.
(b) Mean Xa and IIa concentration profiles at the channel outlet; magenta lines indicate $t^*_{\mathrm{max}}$.
(c–d) Normalized IIa concentration scatter plots at $t^*_{\mathrm{peak}}$ for Pe$=$10 and Pe$=$70, respectively, with matched axis ranges. The SDPD length scale is nondimensionalized by the channel height $H$.
(e–f) IIa concentration histograms: lilac for values at $t^*_{\mathrm{peak}}$ in (c–d), red for values at $t^*_{\mathrm{max}}$.}
\label{fig:RD_Ratto_multiplePe1070_Relow}
\end{figure}

Fig. \ref{fig:RD_Ratto_multiplePe1070_Relow}a shows the peak concentration profiles of Factor Xa (blue) and IIa (yellow) for the two Pe values. At lower Pe, Factor Xa stabilizes more rapidly and the global IIa peak ($t^*_{\mathrm{peak}}$) is reduced. These differences become less evident in the outlet-averaged profiles (Fig. \ref{fig:RD_Ratto_multiplePe1070_Relow}b), underscoring that TGCs alone may not fully reflect spatially localized dynamics. The scatterplots (Fig. \ref{fig:RD_Ratto_multiplePe1070_Relow}c–d) show particle-resolved IIa concentrations normalized to the largest peak, that of Pe$=$70, while the histograms (Fig. \ref{fig:RD_Ratto_multiplePe1070_Relow}e–f) report the number of fluid particles binned by IIa concentration. All data are taken at the times of maximal absolute IIa (cf. Fig. \ref{fig:RD_Ratto_multiplePe1070_Relow}a) and peak outlet IIa (cf. Fig. \ref{fig:RD_Ratto_multiplePe1070_Relow}b).
At Pe = 10, IIa forms compact, localized structures (Fig. \ref{fig:RD_Ratto_multiplePe1070_Relow}c) with narrow concentration distributions (Fig. \ref{fig:RD_Ratto_multiplePe1070_Relow}e). At Pe = 70, IIa becomes more elongated and flow-aligned (Fig. \ref{fig:RD_Ratto_multiplePe1070_Relow}d), accompanied by a broader spread of local concentrations (Fig. \ref{fig:RD_Ratto_multiplePe1070_Relow}f). The broader histogram at Pe = 70 (Fig. \ref{fig:RD_Ratto_multiplePe1070_Relow}f) coincides with the global maximum at $t^* \approx 260$ (Fig. \ref{fig:RD_Ratto_multiplePe1070_Relow}a), indicating stronger spatial variability associated with faster transport.

Overall, these results indicate that higher Pe not only enhances advective transport but also alters the spatial distribution of coagulation species, increasing heterogeneity in local IIa concentrations. Such heterogeneity cannot be inferred from TGCs, which provide only global outlet-averaged kinetics. Complementary spatially resolved descriptors are therefore required to fully characterize coagulation in the microvasculature.
\subsubsection{Role of Reynolds number in microvascular coagulation}\label{subsubsec:Resultsinvivo-Revary}
We next examined the influence of Re on coagulation dynamics under microvascular conditions. Simulations were performed at Pe $= 100$, with Re $= 0.02$, 0.08, 0.1 and 0.5, corresponding to diffusion–advection balances relevant to arterioles and venules (see Appendix \ref{app:subsec:RePe}).

\begin{figure}[!htbp]
\centering
\includegraphics[width=.67
\linewidth]{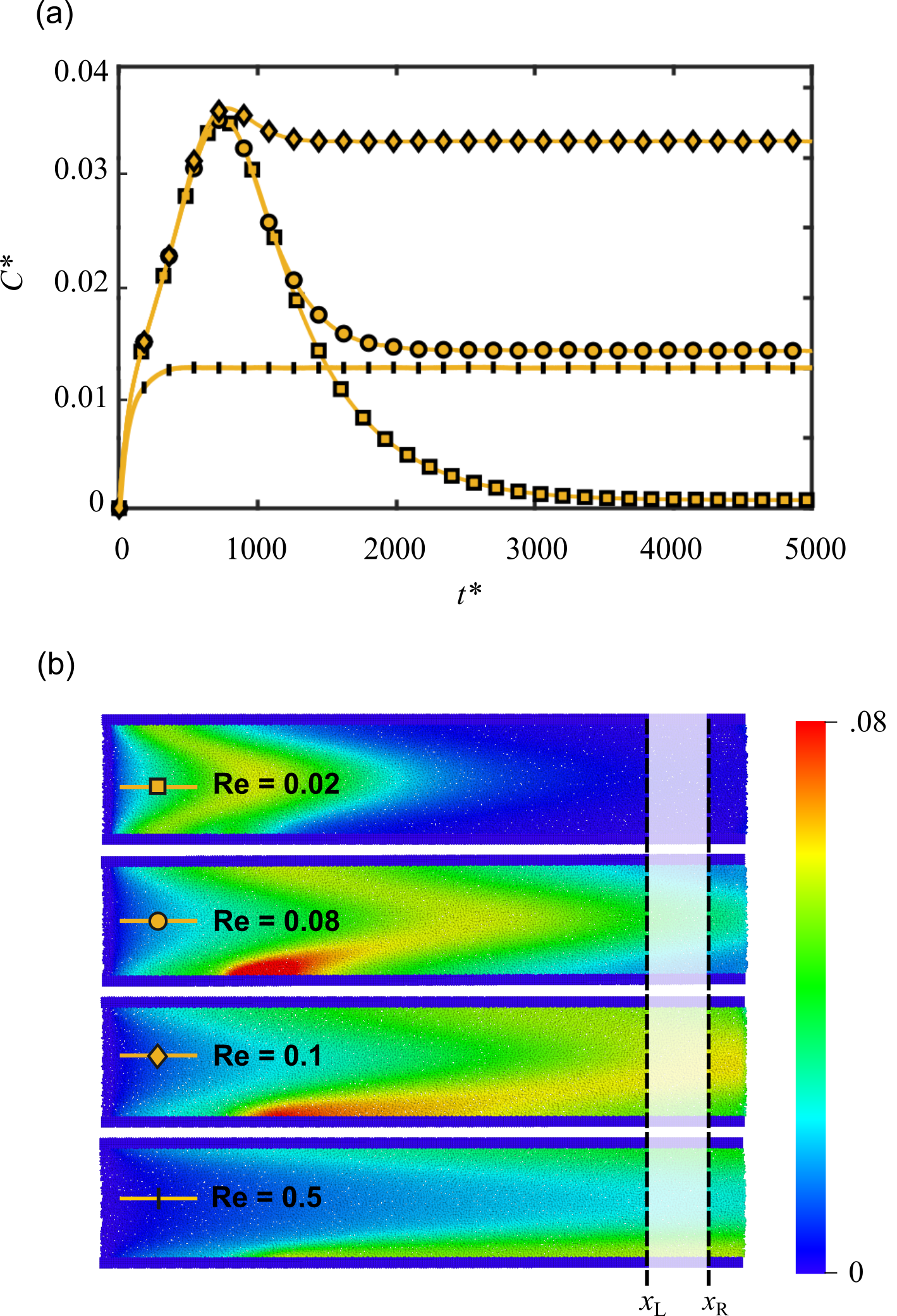}
\caption{(a) Evolution of thrombin (IIa) concentration at the channel outlet for different Reynolds numbers (Re$=$0.02, 0.08, 0.1, and 0.5) with Pe$=$100. Markers indicate Re$=$0.02 (square), Re$=$0.08 (circle), Re$=$0.1 (diamond) and Re$=$0.5 ($|$).  
(b) Spatial distribution of IIa in the \textit{in vivo} domain at $t^* =$ 5000 for the each Re value, fading the channel outlet $x$-slab where local averages are computed. A consistent rainbow color scale is used across all cases for direct comparison.} 
\label{fig:RD_Ratto_multipleRe_Pe100}
\end{figure}

The outlet-averaged IIa profiles (Fig. \ref{fig:RD_Ratto_multipleRe_Pe100}a) show a sharp rise followed by a decay and plateau, except for the highest Re (0.5, $|$), where IIa reaches a steady value without a pronounced peak. These temporal trends are consistent with the spatial maps at $t^* = 5000$ (Fig. \ref{fig:RD_Ratto_multipleRe_Pe100}b), where the faded regions at the channel outlet indicate the slabs used for spatial averaging, corresponding to the plateau concentrations shown in Fig. \ref{fig:RD_Ratto_multipleRe_Pe100}a.

The maps reveal that IIa remains concentrated near the bottom-wall injury at intermediate Re (0.08 and 0.1, circles and diamonds), consistent with a stronger convective washout with larger Re, and reflected at the larger magnitude for mean profiles in the plateaus of Fig. \ref{fig:RD_Ratto_multipleRe_Pe100}a, while at Re = 0.5, stronger advection enhances downstream transport but limits local accumulation intensity near the injury. Across all cases, however, the concentration front extends downstream and arches upward, reaching the upper wall despite the absence of any source there. This apparent symmetry results from the left-to-right inflow of reactive species and their rapid vertical homogenization, driven by the combined effects of diffusion and flow. 
The observed arch-like pattern indicates a regime with high Da $\gg$ 1. Computed timescales (Appendix Table \ref{tab:Timescales_Dimensionless}) give $\mathrm{Da}^{\mathrm{IIa}}\approx58$, confirming a reaction-dominated regime in our parameter window, where fast reaction kinetics further promote homogenization across the channel height. These findings are consistent with previous microfluidic studies under Re$<$1, which also reported laminar, reaction-dominated conditions \cite{tsai2011vitro,schoeman2017,zhao2021hemodynamic}. Likewise, confinement effects have been shown to modulate activation thresholds in coagulation networks \cite{shen2009confinement}, supporting our interpretation that in microvascular-like channels, subtle variations in Re, Pe, or aspect ratio can shift the balance between localized activation and homogenized downstream propagation.

\subsubsection{Impact of injury size}\label{subsubsec:Resultsinvivo-Capsize}

We simulated three cap sizes in a channel with radii $r=1.5$, $4.5$, and $8.5$, corresponding to $r/H=0.15$, $0.45$, and $0.85$ (denoted S, M, and L). 

\begin{figure}[!htbp]
\centering
\includegraphics[width=.99\linewidth]{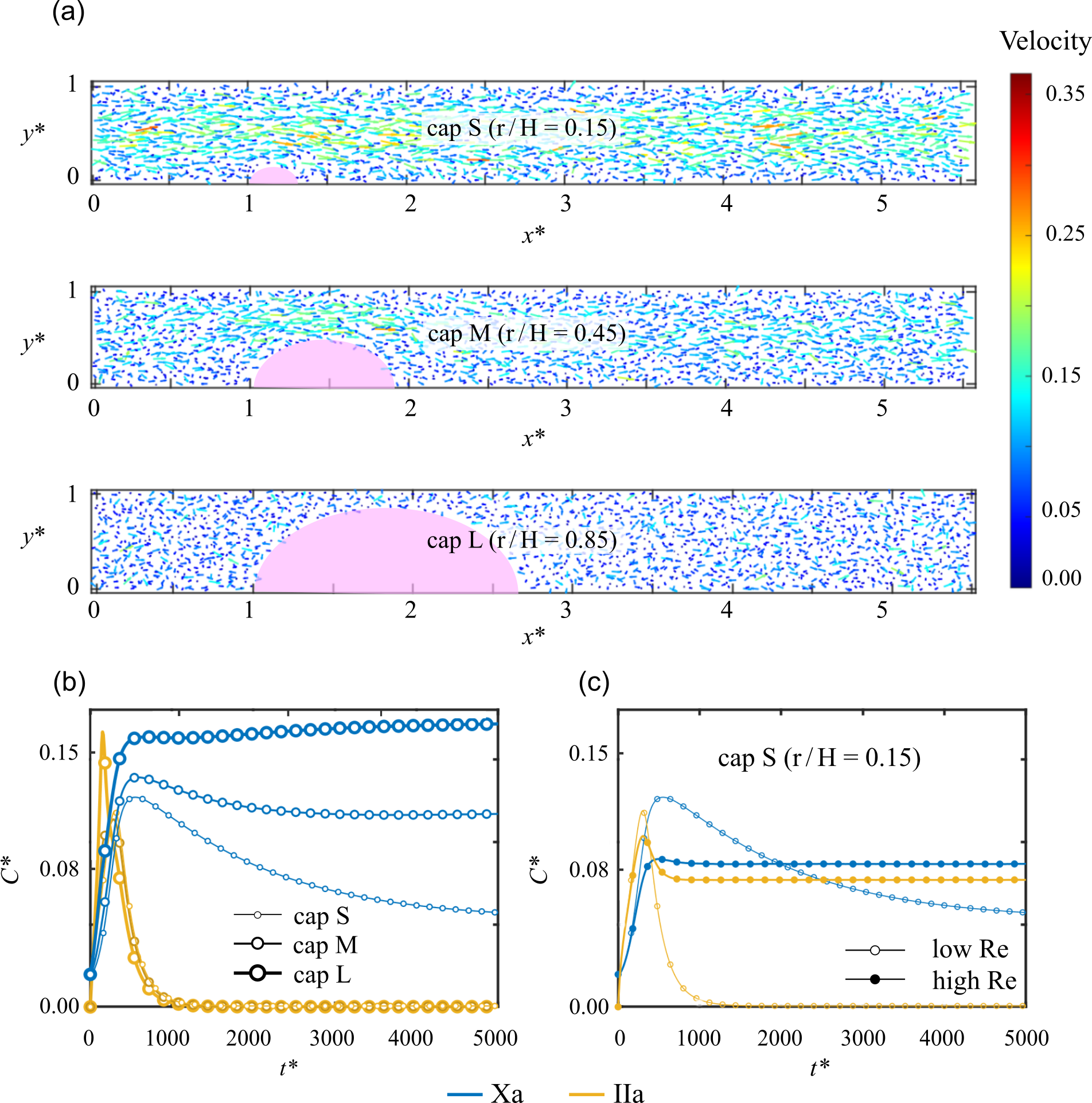}
\caption{Impact of injury size on coagulation dynamics across three cap sizes. (a) Particle velocity fields at the end of the simulation for small (S), medium (M), and large (L) cap sizes. (b) Mean concentration profiles of Factor Xa (blue) and thrombin (yellow) at the channel outlet for each cap size under low Re (0.02). (c) Comparison of mean concentration profiles of Xa and thrombin for the small cap (S) with varying Re: low (0.02) vs. high (0.37). The results highlight the effects of injury size and flow conditions on coagulation dynamics.}
\label{fig:RD_Ratto_3Capsizes}
\end{figure}

The velocity fields at $t^*=5000$ (Fig. \ref{fig:RD_Ratto_3Capsizes}a) illustrate how cap size alters hydrodynamics: small caps permit faster flow, whereas the largest cap (L) markedly reduces it. Outlet profiles of Xa and IIa at Re$=$0.02 (Fig. \ref{fig:RD_Ratto_3Capsizes}b) confirm that larger caps sustain higher Xa fluxes and, in turn, stronger and more stable thrombin responses. IIa peaks early (before 500s) in all cases but persists most effectively when the cap is large. Finally, probing the small cap (S) across flow regimes (Fig. \ref{fig:RD_Ratto_3Capsizes}c) reveals that high Re (0.37) helps maintain stable IIa levels after an initial peak, as in Subsection \ref{subsubsec:Resultsinvivo-Revary}, while at low Re (0.02) IIa declines toward zero. Xa dynamics mirror these trends, with sharper but less sustained responses at lower Re.

Overall, these results show that cap size governs severity by combining geometric and biochemical effects: a larger procoagulant surface both slows local flow and increases Xa flux, amplifying thrombin generation.

It is worth noting that all capped configurations yield IIa levels 50$\%$ higher than in the uncapped case. This occurs because the cap provides a concentrated, sustained Xa source that drives IIa formation more efficiently than the diffuse injury, representing a more localized and severe injury regime. In this context, the absolute magnitude of IIa is not our main focus; rather, the key point is how geometric confinement and surface-mediated fluxes amplify the coagulation response.

Overall, these results demonstrate that injury size and flow conditions strongly modulate coagulation dynamics: larger injuries sustain Xa activation longer, whereas higher flow helps maintain stable concentrations.

\subsubsection{Impact of fibrinogen levels on pathological clot formation}
\label{subsubsec:Resultsinvivo-Fbn}
We simulated the full intrinsic coagulation cascade (Eqs. \eqref{eq:RD_Ratto}) to examine how fibrinogen influences thrombin-mediated fibrin formation. No advection was imposed in these simulations, isolating the effects of diffusion and biochemical reactions.

\begin{figure}[!htbp]
\centering
\includegraphics[width=.79\linewidth]{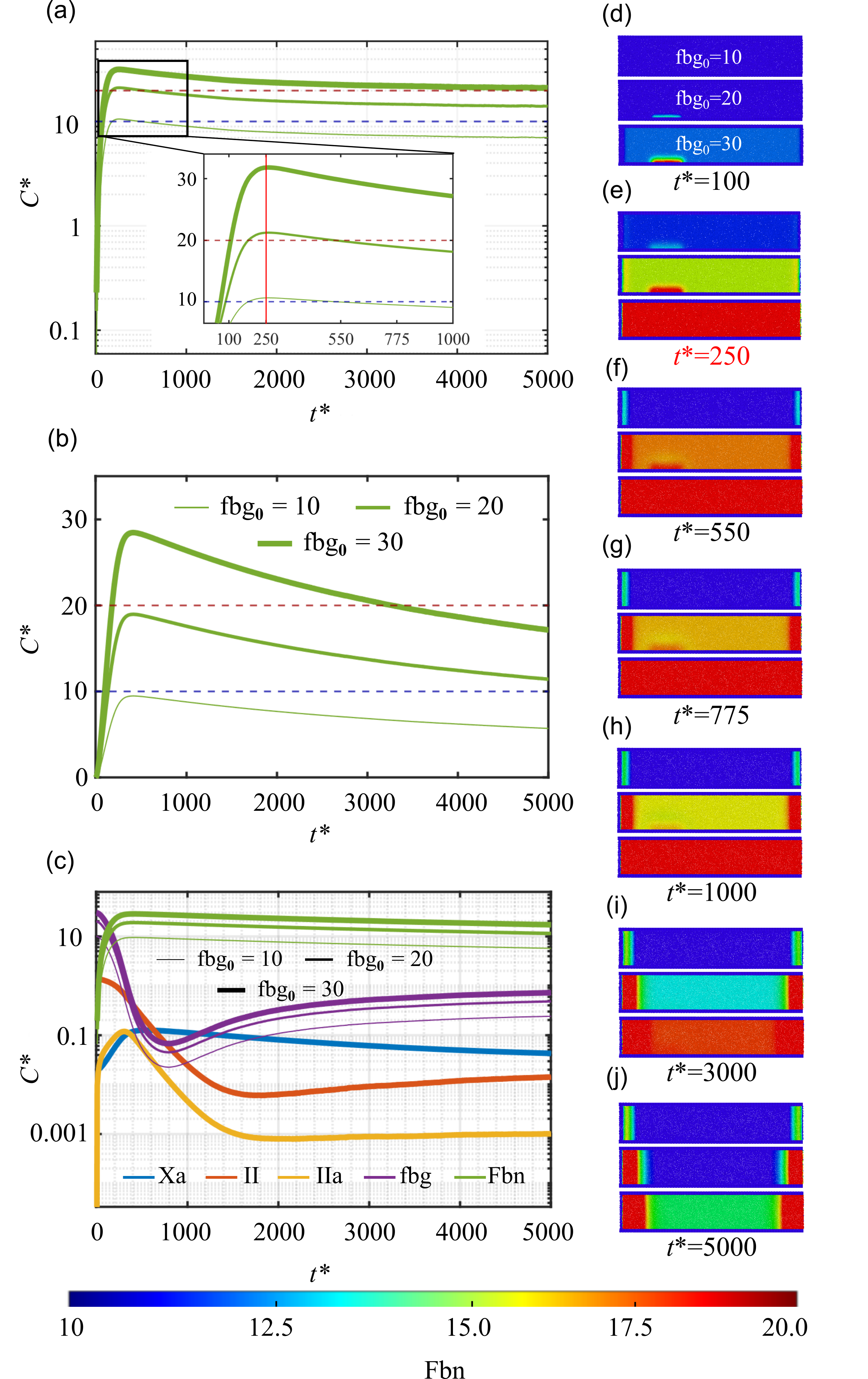}
\caption{Impact of initial fibrinogen concentration ($\text{fbg}_0$) on fibrin (Fbn) dynamics.  
(a) Absolute maximum fibrin (Fbn) concentration over all fluid particles, shown on a logarithmic scale. Dashed horizontal lines mark the physiological (10) and pathological (20) Fbn thresholds. The inset expands the early-time dynamics ($t^* \leq$1000), highlighting the $t^* =$ 250 time point that corresponds to the spatial profiles shown in subfigure (e).
(b) Mean Fbn concentration at the channel outlet for three input levels $\text{fbg}_0=$ 10, 20 and 30). 
(c) Mean outlet concentration profiles for all simulated species, showing that only fbg and Fbn vary with input levels, while the others remain unchanged.
(d–j) Spatial distribution of Fbn at selected time points.}
\label{fig:RD_Ratto_fbg0Explor}
\end{figure}

Higher inputs drive sustained fibrin (Fbn) deposition above pathological thresholds ($\gtrsim 20$), while physiological levels yield only transient, low-amplitude accumulation (Fig.~\ref{fig:RD_Ratto_fbg0Explor}a).
Outlet concentrations remain below pathological levels only for $\mathrm{fbg}_0 = 10$, reflecting limited downstream propagation (Fig.~\ref{fig:RD_Ratto_fbg0Explor}b). Varying fbg$^0$ primarily affects its own concentration and that of Fbn, with minimal impact on IIa or other species (Fig.~\ref{fig:RD_Ratto_fbg0Explor}c). Spatial maps (Fig.~\ref{fig:RD_Ratto_fbg0Explor}d-j) further illustrate pronounced Fbn accumulation throughout the channel for higher fbg inputs. At the representative time point $t^* = 250$, Fig.~\ref{fig:RD_Ratto_fbg0Explor}e) shows the spatial Fbn profiles for the three initial fibrinogen levels. The $\mathrm{fbg}_0 = 30$ case displays a sharply localized, high-amplitude peak characteristic of a severe pro-thrombotic state; $\mathrm{fbg}_0 = 20$ yields a moderate, spatially confined rise consistent with intermediate risk; and $\mathrm{fbg}_0=10$ remains below pathological levels across the domain, reflecting a physiologically safe regime.

Both IIa and Fbn operate in a reaction-dominated regime ($\mathrm{Da}\gg1$), consistent with the rapid vertical homogenization observed in the spatial maps. While IIa exhibits moderate dominance of reaction over diffusion ($\mathrm{Da}\approx60$), Fbn kinetics are substantially faster ($\mathrm{Da}\approx9\times10^3$), explaining its prompt local stabilization once thrombin becomes available. See Appendix Table \ref{tab:Timescales_Dimensionless}.

Consequently, varying fbg input does not alter the transport–reaction balance but shifts the available substrate pool, leading to proportionally greater Fbn formation. This interpretation is consistent with the negligible variation observed in species not directly coupled to fbg conversion and aligns with experimental evidence that clot growth and density scale with fbg availability \cite{davalos2012fbg,wolberg2023fbg}.

\subsection{\textit{In vitro} results}\label{subsec:Resultsinvitro}

We next simulate a cuvette-like \textit{in silico} setup inspired by the Thrombodynamics (TD) assay \cite{sinauridze2018thrombodynamics} to examine fibrinogen’s influence on clot growth. TD is a relevant benchmark because it imposes a localized, immobilized source of tissue factor (TF), creating a spatially asymmetric initiation site similar to a damaged endothelial patch. This preserves the experimental control of an \textit{in vitro} assay while introducing a physiologically meaningful trigger that drives a reaction–diffusion front.

In our simulations, this localized TF source is implemented as an immobile region in the upper layer of the cuvette domain (Fig.~\ref{fig:ADR_domains}b). To assess how fibrinogen availability shapes clot development, we performed simulations with initial fibrinogen concentrations ($\mathrm{fbg}_0$) of 1, 10, 20, and 30. The values 10, 20, and 30 correspond to physiological and elevated fbg levels \cite{wolbergsang2022,davalos2012fbg,kattula2017fbg} and have been used in the preceding results subsection as well to probe model behaviour across clinically relevant conditions. The additional $\mathrm{fbg}_0 = 1$ case serves as an extreme low-fbg scenario, included to highlight sensitivity limits and to contrast with the higher-substrate regimes.

\begin{figure}[H]
\centering
\includegraphics[width=.7\linewidth]{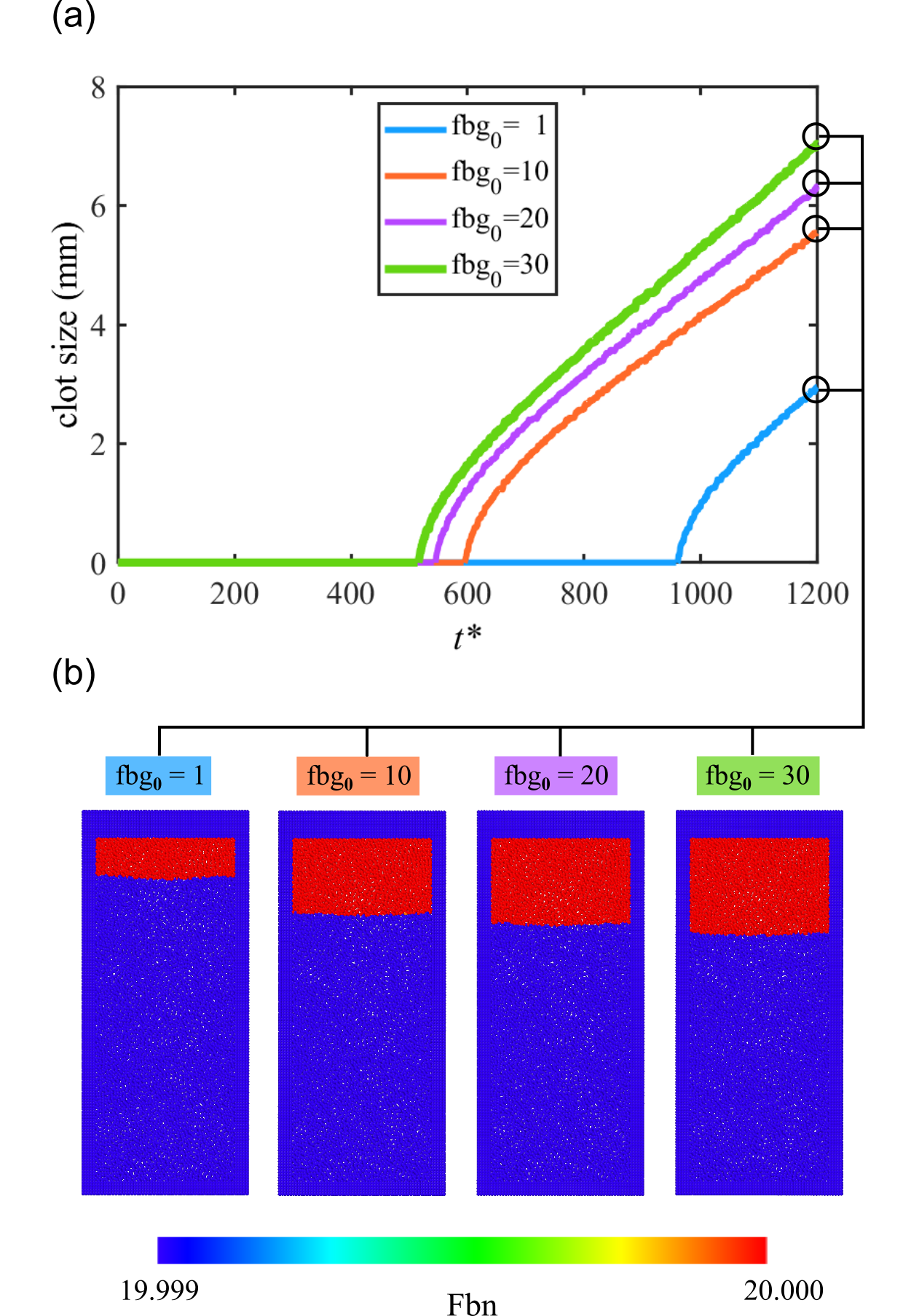}
\caption{\label{fig:RD_csCuvette} (a) Clot size (mm) as a function of dimensionless time $t^*
$, defined by the furthest $y$-coordinate of fluid particles where the normalized Fbn concentration ($C^{\mathrm{Fbn}}$) reaches 20, for four initial fbg levels ($C^{\mathrm{fbg}_0} = 1, 10, 20,$ and $30$).
(b) Final spatial distribution of $C^{\mathrm{Fbn}}$ at $t^* = 1200$ for the same four initial fbg levels.}
\end{figure}

Clot size is defined as the $y$-position (in mm) where Fbn reaches $C^* =$ 20. At low $\mathrm{fbg}_0$, clot formation is markedly delayed ($t^* \sim 1000$), whereas higher $\mathrm{fbg}_0$ produces rapid and sustained Fbn deposition (Fig.~\ref{fig:RD_csCuvette}a–b). These trends indicate that fbg availability regulates both polymerization and effective diffusion: increasing $\mathrm{fbg}_0$ promotes strong, localized Fbn buildup while simultaneously reducing mobility, consistent with experimental observations \cite{kotlarchyk2011}. We extended the simulations to $t^* =$ 1200, where Fbn continued to accumulate and eventually caused numerical overflow. These late-time results should be interpreted carefully: beyond $t^* \approx$ 800, outside the window in which the model was originally benchmarked, the reduced system may fall outside its valid operating regime, as it lacks experimentally calibrated mechanisms that would regulate long-term Fbn growth.

Overall, even under negligible advection, the simulations demonstrate the dominant role of fibrinogen in governing clot growth kinetics and spatial extent. This highlights the ability of the SDPD framework to capture diffusion-limited, spatially resolved coagulation dynamics.

\section{Conclusion}\label{sec13}
Using a fluctuating-hydrodynamics particle-based framework \cite{echeverria2025smoothed}, we examined how the interplay between transport and biochemical kinetics shapes the early stages of thrombin generation in microvascular environments. By embedding validated reduced coagulation schemes into the SDPD method, our aim was to understand how flow strength, dispersion, and injury geometry modulate the spatial organisation and temporal progression of clotting reactions.

Our findings reveal that transport–reaction coupling exerts a dominant influence on coagulation dynamics. Changes in the Péclet number modify not only the rate of transport but also the spatial distribution of thrombin, with heterogeneities that cannot be inferred from outlet-averaged metrics such as TGCs. Variations in Reynolds number (Re$=$0.02–0.5) further reshape dispersion patterns: low Re confines thrombin to the injury vicinity, intermediate Re promotes wall-aligned accumulation, and higher Re attenuates thrombin amplitudes. Injury geometry introduces an additional layer of regulation: capped configurations with concentrated surface-mediated Xa fluxes markedly amplify thrombin formation relative to diffuse surface injuries, demonstrating how geometric confinement can enhance coagulation severity. Complementary cuvette-like simulations reproduced key qualitative features of the Thrombodynamics assay \cite{sinauridze2018thrombodynamics}, including the sensitivity of clot propagation to fibrinogen concentration \cite{wolberg2023fbg,kotlarchyk2011}, underscoring the flexibility of the framework.

Taken together, these results emphasize that the behaviour of reduced coagulation networks can change substantially once embedded in a spatially resolved flow environment. Capturing the transition between diffusion-dominated and advection-enhanced regimes is therefore essential for evaluating coagulation mechanisms in physiologically relevant settings. Future extensions incorporating thrombus permeability, fibrin mechanics, and cell–cell interactions will help connect molecular-scale reactions networks and emergent clot architecture and rheology \cite{fogelson2015, zohravi2025}. With targeted validation in microfluidic systems, the present mesoscale approach offers a basis for predictive, multiscale, and ultimately patient-specific models of clot formation.

\backmatter








\section*{Declarations}
\bigskip





\subsection*{Funding}
We acknowledge funding by the Basque Government through the BERC 2022-2025 program and by the Ministry of Science, Innovation and Universities: BCAM Severo Ochoa accreditation CEX2021-001142S/MICIN/AEI/10.13039/501100011033.

\subsection*{Competing interests}
The authors have no competing interests to declare that are relevant to the content of this article.

\subsection*{Ethics approval and consent to participate}
Not applicable.

\subsection*{Consent for publication}
All authors have read and approved the final manuscript and consent to its publication.

\subsection*{Data availability}
This study did not generate new experimental or observational datasets. The coagulation reaction networks, kinetic parameters, and ODE formulations used in the simulations are drawn entirely from previously published models. All inputs are publicly available in the cited articles.

\subsection*{Materials availability}
Not applicable.

\subsection*{Code availability}
The custom simulation and analysis codes used in this study are available from the corresponding author upon reasonable request.

\subsection*{Author Contributions}
M.E.F. and N.M. designed the research project. M.E.F. performed the simulations and analysed the resulting data. N.M. developed the software and implemented the open-source plugins. M.E. supervised the project and provided overall guidance throughout the study. M.E.F. and N.M. jointly drafted the original version of the manuscript, and all authors contributed to revisions. All authors discussed the results, provided critical feedback, and approved the final manuscript.

\begin{appendices}

\section{Overview of the SDPD formulation}
\label{app:sec:SDPD}
Smoothed Dissipative Particle Dynamics (SDPD) is a mesoscale method that combines the Lagrangian character of particle-based schemes with thermodynamically consistent fluctuating hydrodynamics \cite{Espanol2003,Ellero2003}.
In SDPD, the fluid is represented by particles that carry mass, momentum, and energy, while the conservation laws are recovered through kernel-weighted interactions.

\subsection{Governing equations}\label{app:subsec:SDPD}

The fluid is represented using smoothed Dissipative Particle Dynamics (SDPD), where each particle carries mass, momentum, and compositional fields. The continuum limit of the SDPD model corresponds to the compressible Navier–Stokes equations:
\begin{equation}
\frac{\mathrm{d} \rho}{\mathrm{d} t} = - \rho \nabla \cdot \mathbf{v},
\end{equation}
\begin{equation}
\rho \frac{\mathrm{d} \mathbf{v}}{\mathrm{d} t} + \nabla p - \eta \nabla^2 \mathbf{v} - \left(\zeta + \frac{\eta}{\mathcal{D}}\right) \nabla (\nabla \cdot \mathbf{v}) = 0, \quad \mathcal{D}=2,3,
\end{equation}
where $\rho$ is the density, $\mathbf{v}$ the velocity, $p$ the pressure, $\eta$ the shear viscosity, $\zeta$ the bulk viscosity, and $\mathcal{D}$ the spatial dimension.

At the particle level, the momentum equation includes reversible pressure forces, viscous dissipation, and stochastic contributions:
\begin{align}
m \frac{\mathrm{d} \mathbf{v}_i}{\mathrm{d} t} &= 
- \sum_j \Bigg[ \frac{p_i}{d_i^2} + \frac{p_j}{d_j^2} \Bigg] F_{ij} \mathbf{r}_{ij} 
- \sum_j \Big[ a \mathbf{v}_{ij} + b (\mathbf{v}_{ij} \cdot \mathbf{e}_{ij}) \mathbf{e}_{ij} \Big] \frac{F_{ij}}{d_i d_j} \nonumber\\
& \quad + \sum_j \Big( A_{ij} \, \mathrm{d}\overline{\mathbf{W}}_{ij} + B_{ij} \frac{1}{D} \operatorname{tr}[\mathrm{d} \mathbf{W}_{ij}] \Big) \cdot \frac{\mathbf{e}_{ij}}{\mathrm{d} t}.
\end{align}
where $m$ is the particle mass, $\mathbf{r}_{ij}$ the separation vector, $\mathbf{v}_{ij}$ the relative velocity, $F_{ij}$ the derivative of the smoothing kernel, $\mathbf{e}_{ij}$ the unit vector along $\mathbf{r}_{ij}$, and $A_{ij}, B_{ij}$ weight the stochastic contributions $\mathrm{d}\mathbf{W}_{ij}$.

Species concentrations are carried by each particle and evolved via an advection–diffusion–reaction equation:
\begin{equation}
\frac{\mathrm{d} C^\alpha}{\mathrm{d} t} = - C^\alpha (\nabla \cdot \mathbf{v}) + \nabla \cdot \frac{T C^\alpha}{\xi^\alpha} \nabla \frac{\mu^\alpha}{T} + S^\alpha,
\end{equation}
where $C^\alpha$ is the concentration of species $\alpha$, $T$ the temperature, $\xi^\alpha$ the friction coefficient, $\mu^\alpha$ the chemical potential, and $S^\alpha$ represents source or sink terms, including reactions and localized fluxes. This formulation ensures coupled transport and reaction dynamics are resolved consistently with the underlying fluid motion.

The discretised composition evolution of a blood constituent $\alpha$, in a particle $i$, is computed using the traditional SDPD interpolation, with the neighboring $j$ particles, and is given by 
\begin{equation}
    \frac{\der C_i^{\alpha}}{\der t} = \sum_j D_{ij}^{\alpha} C_{ij}^{\alpha}\frac{F_{ij}}{d_i d_j} + S_i^{\alpha},
    \label{eq:compBalance_discrete} 
\end{equation}
with pairwise diffusivity $D_{ij}^\alpha = 2 D^\alpha d_i d_j / (d_i + d_j)$ and kernel derivative $F_{ij}$ evaluated from the SDPD smoothing kernel $W_{ij}$. This pairwise form preserves symmetry, conservation, and thermodynamic consistency, and reduces to Fickian diffusion in the dilute limit. Further theoretical background and a derivation from the GENERIC framework can be found in \cite{Espanol2003,ellero2018everything}.

In the simulations of this study, the fluid was discretized using particles of spacing $\Delta x = 0.2$ and kernel cutoff $h = 4\Delta x$. Time integration used a stochastic velocity–Verlet scheme with timestep ($\Delta t = 10^{-4},\mathrm{s}$. Concentration fields for plasma species were advanced concurrently using the reaction and diffusion parameters in Tables~\ref{tab:RD_Ratto_modelparameters} and \ref{tab:Chendi-reaction_rates}. A complete summary of SDPD units, material parameters, and numerical constants is provided in Table~\ref{tab:sdpd-params}.


\begin{table}[htbp]
\centering
\caption{\label{tab:sdpd-params}Units, simulation parameters, characteristic timescales, and dimensionless numbers used in the SDPD models.}

\newlength{\tablewidth}
\setlength{\tablewidth}{0.9\linewidth}

\begin{tabular*}{\tablewidth}{@{\extracolsep{\fill}}ll@{}}
\toprule
\multicolumn{2}{c}{\textbf{Basic Units}} \\
\midrule
Length & $\ulen$ \\
Energy & $k_B T$ \\
Mass & $m_f$ \\
\bottomrule
\end{tabular*}

\vspace{.33em}

\begin{tabular*}{\tablewidth}{@{\extracolsep{\fill}}ll@{}}
\toprule
\multicolumn{2}{c}{\textbf{Derived Units}} \\
\midrule
Time & $\tau = \ulen \sqrt{m_f / (k_B T)}$ \\
Velocity & $u = \ulen / \tau = \sqrt{k_B T / m_f}$ \\
Diffusion constant & $D = \ulen^2 / t_0 = \ulen \sqrt{k_B T / m_f}$ \\
Dynamic viscosity & $\eta = m_f / (t_0 \ulen) = \sqrt{m_f k_B T} / \ulen^2$ \\
Kinematic viscosity & $\nu = \ulen^2 / t_0 = \ulen \sqrt{k_B T / m_f}$ \\
\bottomrule
\end{tabular*}

\vspace{.33em}

\begin{tabular*}{\tablewidth}{@{\extracolsep{\fill}}lc@{}}
\toprule
\textbf{Parameter} & \textbf{Value (non-dimensional)} \\
\midrule
Mass density, $\rho$ & 1.0\\
Dynamic viscosity, $\eta$ & 3.0 \\
Kinematic viscosity, $\nu$ & 3.0 \\
Thermal energy, $k_B T$ & 0 \\
Speed of sound, $c_s$ & 10 \\
Time step, $\Delta \tau$ & $10^{-3}\,\tau$ \\
Reference length, $l$ & 1 (unit length) \\
Reference time, $\tau$ & 1 (unit time) \\
Simulation box length & $L_x$ (varied)\\
Discretization spacing, $dx$ & 0.2 \\
Cutoff radius, $h$ & 4\,$dx$ \\
\bottomrule
\end{tabular*}

\vspace{.33em}

\begin{tabular*}{\tablewidth}{@{\extracolsep{\fill}}ll@{}}
\toprule
\multicolumn{2}{c}{\textbf{Characteristic Timescales and Dimensionless Numbers}} \\
\midrule
Diffusion timescale & $\tau_\mathrm{diff} = {H^2}/{D}$ \\
Advection timescale & $\tau_\mathrm{adv} = {H}/{U}$ \\
Reaction timescale & $\tau_\mathrm{react} = {1}/{k_\mathrm{eff} C_\mathrm{ref}}$ \\
Péclet number & $\mathrm{Pe} = {U H}/{D} = \dfrac{\tau_\mathrm{diff}}{\tau_\mathrm{adv}}$ \\
Reynolds number & $\mathrm{Re} = {\rho U H}/{\eta}$ \\
Damköhler number & $\mathrm{Da} = {\tau_\mathrm{adv}}/{\tau_\mathrm{react}} = k_\mathrm{eff} C_\mathrm{ref} {H}/{U}$ \\
Schmidt number & $\mathrm{Sc} = {\nu}/{D}$ \\
Mach number & $\mathrm{Ma} = {U}/{c_s}$ \\
\bottomrule
\end{tabular*}
\end{table}

\subsection{Transport properties}\label{app:sdpd:properties}
Macroscopic viscosity and diffusivity emerge naturally from the microscopic interaction parameters and kernel choice. The flow is driven by applying a uniform body force per particle, expressed as

\begin{equation}
f_b = \frac{8 \eta v}{\left(1/dh^2\right) H^2},\label{eq:bodyforce}
\end{equation}

where $\eta$ is the fluid viscosity, $v$ is the mean flow velocity, $H$ is the channel half-height, and $dh$ the discretization spacing. This formulation enforces a steady Poiseuille-like profile consistent with the chosen Reynolds and Péclet numbers.

Importantly, the maximum admissible velocity is constrained by the fixed speed of sound $c_s$ in the SDPD fluid, since the Mach number $Ma = v / c_s \ll 1$ must hold to suppress compressibility artifacts. Consequently, Pe and Re values were restricted to ranges consistent with $Ma \lesssim 0.1$, avoiding unphysically large velocities even if higher Pe could be physiologically relevant.


\section{Characteristic timescales and dimensionless numbers}\label{app:sec:ReynoldsPeclet}

To assess the transport and reaction regimes in the simulations, characteristic diffusion ($\tau_\mathrm{diff}$), advection ($\tau_\mathrm{adv}$), and reaction ($\tau_\mathrm{react}$) timescales were computed using the dimensionless channel height ($H=$10) and diffusivities consistent with the reference scales ($H_\mathrm{ref}=100~\mu\mathrm{m}$) and ($t_\mathrm{ref}=1~\mathrm{s}$). From these timescales, the corresponding Péclet ($\mathrm{Pe}$) and Damköhler ($\mathrm{Da}$) numbers were derived to quantify the relative importance of transport and biochemical kinetics. The effective rate constants ($k_\mathrm{eff}=k/k_\mathrm{ref}$) were obtained using ($k_\mathrm{ref}=10^{-3}~\mathrm{nM^{-1}s^{-1}}$). Parameters for thrombin (IIa) and fibrin (Fbn) formation were calculated as summarized in Table \ref{tab:Timescales_Dimensionless}.

\begin{table}[htbp]
\centering
\caption{Characteristic timescales and nondimensional numbers used to describe transport and reaction regimes in the simulated channel. All quantities are expressed in dimensionless simulation units, with $H=10$, $L=55$, and $\mathrm{Pe}=100$.}
\label{tab:Timescales_Dimensionless}
\begin{tabular}{lcccccc}
\toprule
Species & $D^\alpha$ & ${k}^\alpha$ & $\tau_{\text{diff}} = H^2/D^\alpha$ & $\tau_{\text{adv}} = L/U$ & $\tau_{\text{react}} = 1/{k}^\alpha$ & $\mathrm{\mathbf{Da}} = \mathbf{\tau}_{\text{diff}} / \tau_{\text{react}}$ \\
\midrule
Thrombin (IIa) & 0.00647 & $3.76\times10^{-3}$ & $1.55\times10^{4}$ & $8.50\times10^{2}$ & $2.66\times10^{2}$ & $\mathbf{5.8\times10^{1}}$ \\
Fibrin (Fbn) & 0.00150 & $1.33\times10^{-1}$ & $6.67\times10^{4}$ & $3.67\times10^{3}$ & $7.52$ & $\mathbf{8.9\times10^{3}}$ \\
\bottomrule
\end{tabular}
\end{table}

\noindent
The Damköhler numbers confirm that both IIa and Fbn operate in a reaction-dominated regime ($\mathrm{Da}\gg1$), consistent with the rapid vertical homogenization observed in the spatial maps in Subsection \ref{subsubsec:Resultsinvivo-Fbn}.

\subsection{Estimates for microvascular regimes}\label{app:subsec:RePe}
We report here physiological estimates of the Reynolds and Péclet numbers in the microcirculation, which motivate the choice of nondimensional parameters used in this study. Values are derived from canonical plasma and vessel properties, consistent with experimental measurements \cite{goldsmith1986rheological}.

\paragraph{Re number.}  
Using plasma density $\rho=1.03\times 10^3$ kg/m$^3$ and viscosity $\eta=1.3\times 10^{-3}$ Pa·s ($\nu=1.26\times 10^{-6}$ m$^2$/s), together with velocities $u=10^{-3}$–$2\times 10^{-2}$ m/s and vessel diameters $H=5$–$100~\mu$m \cite{goldsmith1986rheological, popel2005}, we obtain
\[
\text{Re} = \frac{\rho u H}{\eta} \approx 0.008\text{–}1.6,
\]
spanning diffusion-dominated to weakly inertial regimes. These values reflect the microvascular regime (arterioles and venules), where viscous effects dominate and flow remains laminar.  

\paragraph{Pe number.}  
The effective transport of solutes in blood is shaped by both molecular diffusion and shear-enhanced dispersion. Reported effective diffusivities for macromolecules and clotting factors lie in the range $D_{\text{eff}}=(1.3$–$21.8)\times 10^{-10}$ m$^2$/s, depending on hematocrit and shear rate \cite{goldsmith1986rheological,rezaeimoghaddam2022continuum,tang2018shear}. Inserting the same velocity and vessel size scales yields
\[
\text{Pe} = \frac{uH}{D_{\text{eff}}} \approx 5\text{–}1.5\times 10^{4}.
\]
These values indicate predominantly convection-driven transport, with the lower bound corresponding to molecular diffusion and the upper bound to red blood cell–mediated dispersion.  

\paragraph{Implications.}  
These estimates place the microvascular regime firmly within low–Reynolds-number flows ($\text{Re}\lesssim 1$), where inertia is negligible, and moderate-to-high Péclet numbers ($\text{Pe}\sim 10^1{-}10^4$), where advection competes with or dominates over molecular diffusion. Such conditions are characteristic of arterioles and venules with diameters of 5–100 $\mu$m \cite{popel2005}. In vivo, shear-enhanced transport by red blood cells further increases effective diffusivity \cite{kumar2012,schoeman2017}, and comparable ranges are reproduced in microfluidic and flow-based \textit{in vitro} assays \cite{zilberman2017,mangin2021vitro}.

\section{Blood coagulation benchmarks}\label{app:BloodCoagulBenchmarks}

\subsection{Overview of reduced coagulation models}
Before running our simulations, we employed established reduced coagulation models \cite{chen2019reduced, ratto2021patient}, representing either the intrinsic or extrinsic pathways shown in Section \ref{sec1:Intro}. These models reproduce characteristic differences in clot growth across healthy and pathologic conditions, providing a basis for exploring treatment effects.

\subsubsection{Reduced intrinsic coagulation model}\label{app:subsubsec:Ratto}
\begin{figure}[H]
\centering
\includegraphics[width=.58\linewidth]{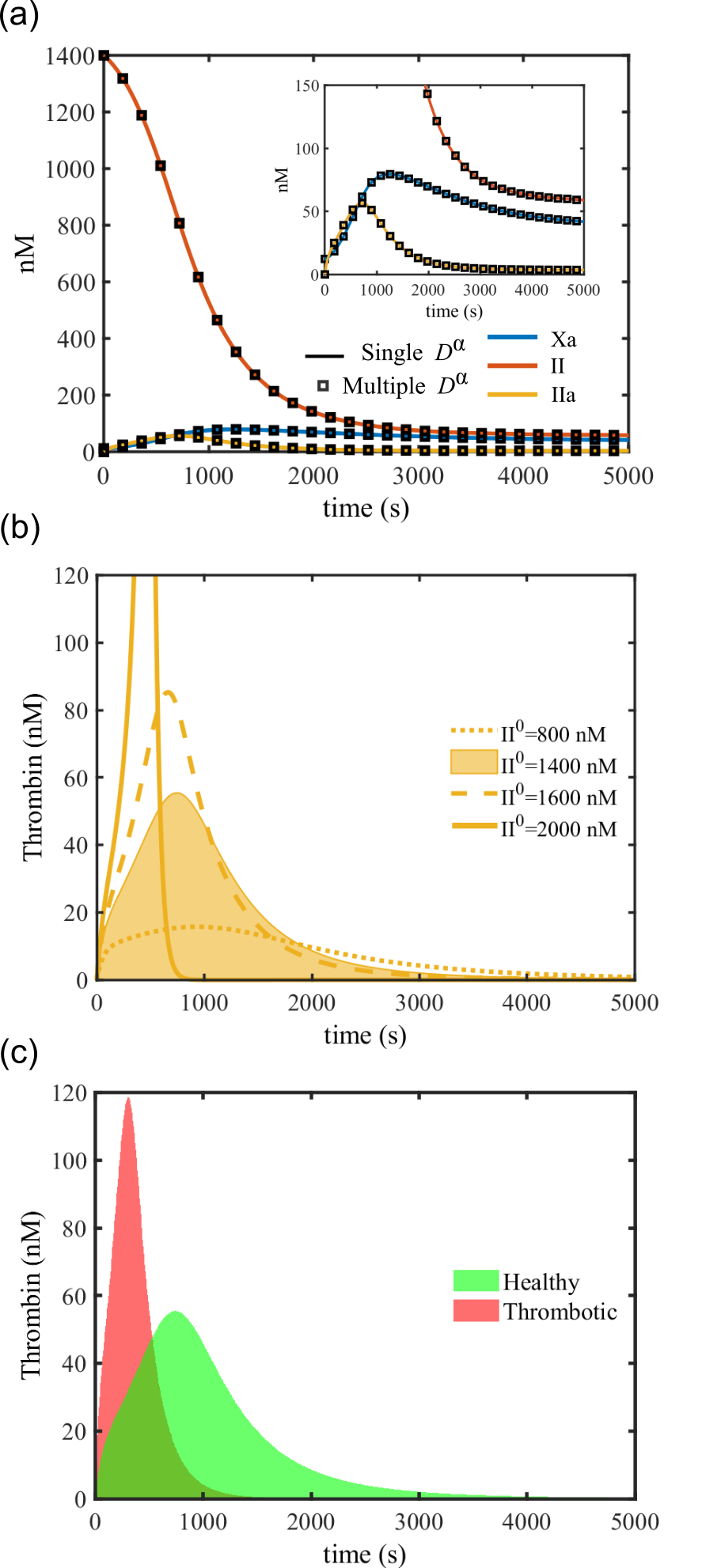}
\caption{Thrombin Generation Curves (TGCs) from Ratto's model under different conditions. (a) Mean concentrations of Xa (blue), II (red), and IIa (yellow) in a reaction-dominant healthy system; square markers indicate species-specific diffusivities ($D^\alpha$), matching uniform diffusivity results. (b) Effect of varying initial prothrombin (II$^{0}$) while keeping healthy reaction rates. (c) Comparison of healthy vs. thrombotic TGCs, showing differences in thrombin peak magnitude and timing.}
\label{fig:RD_Ratto_checkrates_TGCsArea}
\end{figure}

Parameter adjustments in the Ratto model \cite{ratto2021patient} allow calibration to patient-specific thrombin generation curves (TGCs), demonstrating its adaptability. Fig.~\ref{fig:RD_Ratto_checkrates_TGCsArea} summarizes the model behavior under different conditions.
In Fig. \ref{fig:RD_Ratto_checkrates_TGCsArea}a, the main species —Xa, II, and IIa— are shown for a reaction-dominated healthy system. The near overlap between simulations using species-specific diffusivities ($D^\alpha$; square markers) and a uniform diffusivity confirms that such variations have negligible impact on thrombin kinetics, consistent with the diffusion-dominated homogenization seen in channel simulations. Fig. \ref{fig:RD_Ratto_checkrates_TGCsArea}b illustrates how increasing initial prothrombin (II$^0$) accelerates and amplifies thrombin formation, reproducing experimental trends observed in hyperprothrombinemic plasma. Finally, Fig. \ref{fig:RD_Ratto_checkrates_TGCsArea}c compares healthy and thrombotic parameter sets, showing a sharper and earlier thrombin peak in the latter. These results demonstrate how modest parameter changes, particularly in Xa-mediated reactions, can reproduce clinically relevant TGC profiles, highlighting the reduced Ratto model’s ability to capture diverse coagulation phenotypes. Table~\ref{tab:RD_Ratto_modelparameters} lists the parameters.

\begin{table}[ht]
    \centering
    \caption{Values of reaction kinetics, diffusion coefficients, and initial concentrations adopted in the intrinsic coagulation model. The Healthy and Thrombotic columns list values for normal and pathological conditions, respectively, reported by Ratto el al. \cite{ratto2021patient}.}
\begin{tabular}{c|c|c|c|c}
\hline \multicolumn{5}{c}{  \textbf{Biochemical Reactions Kinetic Constants} }\\
\hline \textbf{Symbol} & \textbf{Healthy} & \textbf{Thrombotic} & \textbf{UM} & \textbf{Ref.} \\ 
\hline $k_1$ & $4.082\times10^{-05}$ & $4.082\times10^{-05}$ & s$^{-1}$ & \cite{Ratto2020clustering} \\ 
 $k_2$ & $1.111\times10^{-05}$ & $1.111\times10^{-05}$ & nM$^{-1}$s$^{-1}$ & \cite{Ratto2020clustering} \\
 $k_3$ & $2.4482\times10^{-07}$ & $2.4482\times10^{-07}$ & nM$^{-2}$s$^{-1}$ & \cite{Ratto2020clustering} \\
 $k_4$ & $0.000379$ & $0.000479$ & nM$^{-1}$s$^{-1}$ & \cite{Ratto2020clustering} \\
 $k_5$ & $1.859\times10^{-05}$ & $3.859\times10^{-05}$ & nM$^{-1}$s$^{-1}$ & \cite{Ratto2020clustering} \\
 $k_6$ & $3.76210^{-06}$ & $4.762\times10^{-06}$ & nM$^{-1}$s$^{-1}$ & \cite{Ratto2020clustering} \\
 $k_7$ & $1.285 \times 10^{-10}$ & $1.285 \times 10^{-10}$ & nM$^{-2}$s$^{-1}$ & \cite{Ratto2020clustering} \\
 $k_8$ & $4.066\times10^{-10}$ & $4.066\times10^{-10}$& nM$^{-3}$s$^{-1}$ & \cite{Ratto2020clustering} \\
 $k_9$ & $0.02$ & $0.009$ & nM$^{-1}$s$^{-1}$ & \cite{Ratto2020clustering} \\
 $\alpha_1$ & $1.33\times10^{-04}$ & $1.117\times10^{-03}$ & nM$^{-1}$s$^{-1}$ & \cite{Ratto2020clustering} \\
 $\alpha_2$ & $1.52\times10^{-04}$ & $1.67\times10^{-03}$ & nM$^{-1}$s$^{-1}$ & \cite{Ratto2020clustering} \\
 
 \hline \multicolumn{5}{l}{ \textbf{Diffusion Coefficients} } \\
 \hline $D^{\text{Xa}}$ & \multicolumn{2}{c|}{$7.37\times10^{-07}$} & $\mathrm{~cm}^2 \mathrm{~s}^{-1}$ & \cite{anand2003model} \\
$D^{\text{II}}$ & \multicolumn{2}{c|}{$5.21\times10^{-07}$} & $\mathrm{~cm}^2 \mathrm{~s}^{-1}$ & \cite{anand2003model} \\
$D^{\text{IIa}}$ & \multicolumn{2}{c|}{$6.47\times10^{-07}$} & $\mathrm{~cm}^2 \mathrm{~s}^{-1}$ & \cite{anand2003model} \\
 $D^{\mathrm{fbg}}$ & \multicolumn{2}{c|}{$2.47\times10^{-07}$} & $\mathrm{~cm}^2 \mathrm{~s}^{-1}$ & \cite{anand2003model} \\
 $D^{\text{Fbn}}$ & \multicolumn{2}{c|}{$1.50\times10^{-07}$} & $\mathrm{~cm}^2 \mathrm{~s}^{-1}$ & \cite{anand2003model}\\
  $D^{\text{Fbn}_p}$ & \multicolumn{2}{c|}{$1.00\times10^{-10}$} & $\mathrm{~cm}^2 \mathrm{~s}^{-1}$ & \cite{anand2003model}\\

\hline \multicolumn{5}{l}{ \textbf{Initial Concentrations} } \\
\hline $\mathrm{C}^{\text{X}^0}$ & 135 & 175 & nM & \cite{Ratto2020clustering} \\
 $\mathrm{C}^{\text{II}}$ & 1400 & 1900 & nM & \cite{Ratto2020clustering} \\
 $\mathrm{C}^{\mathrm{fbg}}$ & $\leq$10000 & $>$10000 & nM & \cite{Ratto2020clustering} \\
\hline \multicolumn{5}{l}{ \textbf{Other} } \\
\hline $\gamma$ & \multicolumn{2}{c}{0.007} & . & \\
\hline
\end{tabular}
\label{tab:RD_Ratto_modelparameters}
\end{table}

\subsubsection{Initialization at steady-state}
\label{app:subsubsec:InitSteadyState}

To avoid artificial transients and non-physical reaction delays, all simulations are initialized from steady-state concentrations.  
For Ratto’s model, steady-state values of Factor Xa (\(\mathrm{Xa}\)), Prothrombin (\(\mathrm{II}\)), and Thrombin (\(\mathrm{IIa}\)) are obtained by solving the reaction system~\eqref{eq:RD_Ratto} in Section \ref{subsubsec:intrinsic} with all time derivatives set to zero.

Starting from initial guesses (\(\mathrm{Xa}^0 = 135~\mathrm{nM}\), \(\mathrm{fbg} = 10{,}000~\mathrm{nM}\)) and using the parameter set in Table~\ref{tab:RD_Ratto_modelparameters} for a healthy subject, MATLAB’s \texttt{fsolve} yields the following equilibrium concentrations:
\[
\begin{aligned}
\mathrm{Xa} &\approx 14.99~\mathrm{nM}, \quad
\mathrm{II} \approx 1.11~\mathrm{nM}, \quad
\mathrm{IIa} \approx 0.00~\mathrm{nM},\\[-2pt]
\mathrm{fbg} &\approx 9{,}999.96~\mathrm{nM}, \quad
\mathrm{Fbn} \approx 0.20~\mathrm{nM}, \quad
\mathrm{Fbn_P} \approx 7.66~\mathrm{nM}.
\end{aligned}
\]
These concentrations define the reference steady state for the kinetic parameters considered.


\subsubsection{Reduced extrinsic coagulation model}\label{app:subsubsec:ChenDiEqns}
We use the reduced extrinsic model from Chen and Diamond (2019) \cite{chen2019reduced}. Reaction rates $a_i$ follow Michaelis-Menten kinetics:
$$
a_i = \frac{k_{\text{cat}} [\text{S}]_i}{K_m + [\text{S}]_i},
$$
where $k_{\text{cat}}$ is the catalytic constant, $[\text{S}]_i$ is the substrate concentration, and $K_m$ is the Michaelis constant. Species-specific parameters model physiological conditions and are summarized in Table \ref{tab:Chendi-reaction_rates}.

Authors reported that three effectiveness factors ($\eta_4 = 0.18$, $\eta_5 = 0.05$, and $\eta_6 = 0.36$) needed calibration to match experimental data. The system remains well-defined up to $t = 800$s; beyond this point, the numerical solution diverges and non-finite (NaN) values appear.

\begin{table}[!ht]
\centering
\caption{Parameters for the reduced model of the extrinsic coagulation pathway.}
\label{tab:Chendi-reaction_rates}

\begin{tabular}[t]{c|c|c|c|c|c}
\hline
\textbf{Enzyme} & \textbf{[S]$_0$ ($\mu$M)} & \textbf{$k_{\text{cat}}$ (s$^{-1}$)} & \textbf{$K_m$ ($\mu$M)} & \textbf{$\alpha$ (s$^{-1}$)} & \textbf{$\eta$} \\
\hline
TF/VII    & $\mathrm{X_0}=0.17$   & 1.15    & 0.24  & 0.48~\cite{leiderman2011grow}                       & 1.00 \\
TF/VIIa   & $\mathrm{IX_0}=0.09$  & 1.8     & 0.42  & 0.32~\cite{chatterjee2010systems}                  & 1.00 \\
IXa/VIIIa & $\mathrm{X_0}=0.17$   & 8.2     & 0.082 & 5.53~\cite{chatterjee2010systems}                  & 1.00 \\
Xa/Va     & $\mathrm{II_0}=1.4$   & 30      & 0.3   & 24.7~\cite{chatterjee2010systems}                  & 0.18 \\
IIa       & $\alpha$-fbg = 18     & 80      & 6.5   & 58.8~\cite{chatterjee2010systems,higgins1983steady} & 0.05 \\
IIa/p     & $\mathrm{XI_0}=0.031$ & 0.00013 & 0.05  & $5\times10^{-5}$~\cite{chatterjee2010systems}      & 0.36 \\
XIa/p     & $\mathrm{IX_0}=0.09$  & 0.21    & 0.2   & 0.065~\cite{chatterjee2010systems}                 & 1.00 \\
\hline
\end{tabular}
\vspace{.5em} 
\begin{tabular}[t]{lc}
\hline
\multicolumn{2}{c}{\textbf{Diffusion Coefficients $C^\alpha$} ($\mathrm{cm}^2 \mathrm{s}^{-1}$)} \\
\hline
$D^{\text{TF}}$        & $2.00\times10^{-07}$ \\
$D^{\text{Xa}}$        & $7.37\times10^{-07}$ \\
$D^{\text{IXa}}$       & $7.00\times10^{-07}$ \\
$D^{\text{XIa}}$       & $7.00\times10^{-07}$ \\
$D^{\text{Fbn}}$       & $1.00\times10^{-07}$ \\
$D^{\text{IIa}}$       & $6.47\times10^{-07}$ \\
$D^{E\text{S}}$        & $1.00\times10^{-07}$ \\
$D^{\gamma\text{S}}$   & $1.00\times10^{-07}$ \\
\hline
\end{tabular}
\end{table}
\end{appendices}


\pagebreak
\bibliography{MyCollection}

\end{document}